\documentclass[aps,prl,twocolumn,superscriptaddress,groupedaddress,floatfix,longbibliography]{revtex4-1}
\usepackage[utf8]{inputenc}

\usepackage{graphicx}  
\usepackage{dcolumn}   
\usepackage{bm}        
\usepackage{amssymb}   
\usepackage{amsmath}   
\usepackage{amsthm}    
\usepackage{xcolor}    
\usepackage{tikz}      
\usepackage{ifthen}    
\usepackage{subfigure}
\usepackage{xspace}
\usepackage[toc,page]{appendix}
\newcounter{para}

\begin{document}

\title  {
Melting of generalized Wigner crystals in
transition metal dichalcogenide heterobilayer Moir\'e systems
}

\author{Michael Matty, Eun-Ah Kim}
\email{eun-ah.kim@cornell.edu}

\affiliation{Department of Physics, Cornell University, Ithaca, New York 14853, USA}

\date{\today}

\begin{abstract}
    Moir\'e superlattice systems such as transition metal dichalcogenide heterobilayers
    have garnered significant recent interest due to their promising utility as 
    tunable solid state simulators. 
    Recent experiments on a WSe$_2$/WS$_2$ heterobilayer
    detected incompressible charge ordered states that one can view as generalized Wigner crystals.
    The tunability of the hetero-TMD Moir\'e system presents an opportunity to study the rich set of possible phases
    upon melting these charge-ordered states.
    Here we use Monte Carlo simulations to study these intermediate phases in between incompressible charge-ordered states
    in the strong coupling limit.
    We find two distinct stripe solid states to be each preceded by distinct types of nematic states. 
    In particular, we discover microscopic mechanisms that stabilize each of the nematic states, whose order parameter transforms as the two-dimensional $E$ representation of the Moir\'e lattice point group.
    Our results provide a testable experimental prediction of where both types of nematic occur,
    and elucidate the microscopic mechanism driving their formation.
    
\end{abstract}

\maketitle
\newpage

\begin{figure*}
    \centering
    \includegraphics[width=\linewidth]{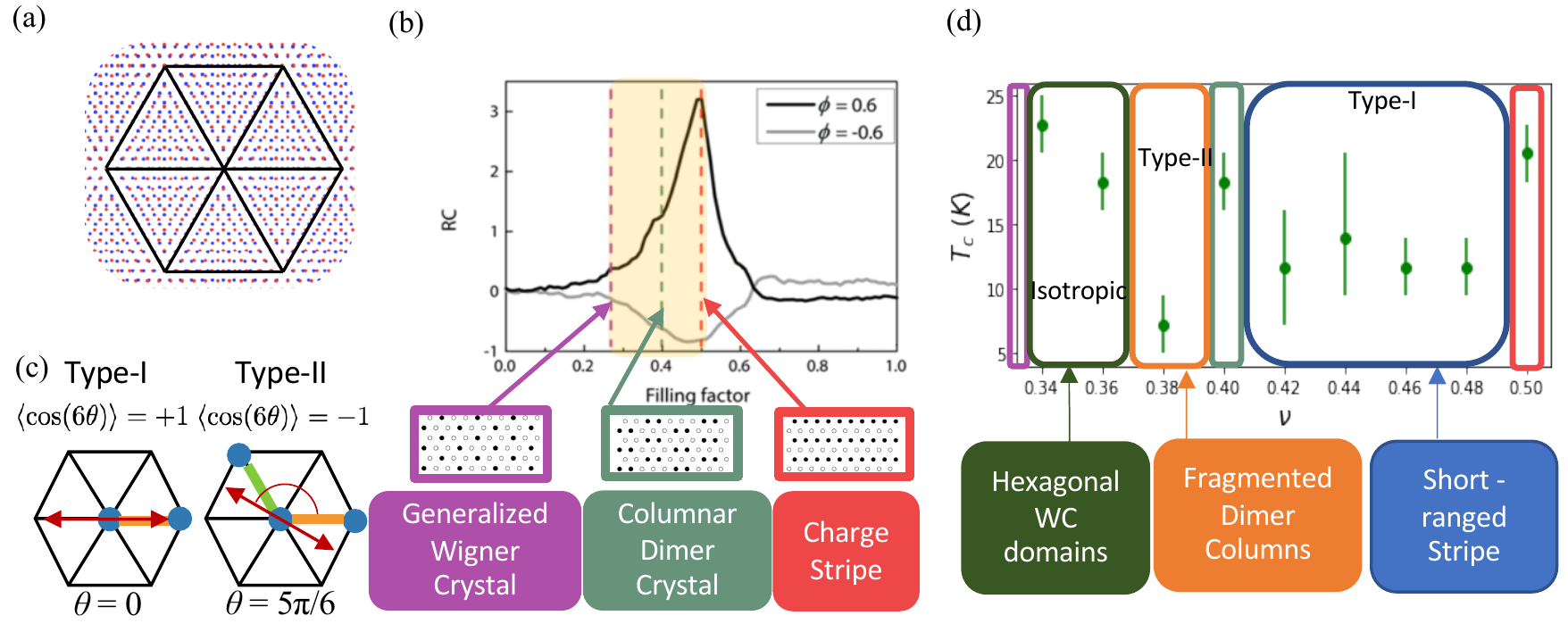}
    \caption{(a) The red and blue dots show the sites of two honeycomb lattices whose
    lattice constants differ by $5\%$. These lattices are layered at zero twist angle,
    resulting in an emergent
    triangular Moir\'e lattice with a unit cell indicated by the black lines.
    In the case of TMD heterobilayers, the Moir\'e lattice has point group $D_3$. 
    (b) Top: optical anisotropy as a function of Moir\'e lattice filling, reproduced
    from ref.~\cite{Jin2021Nat.Mater.}.
    Bottom: charge order patterns at $1/3-,2/5-,$ and $1/2-$ filling as determined
    by Monte Carlo, reproduced from ref.~\cite{Xu2020Nature}.
    (c) On a lattice with $D_3$ symmetry, there are two distinct types of nematic states.
    Type-I nematics (left) have a nematic director oriented along a strong bond orientation 
    at $\theta \in \{0, \pi/3, 2\pi/3 \}$ and have $\langle \cos(6\theta)\rangle = 1$. 
    Type-II nematics (right) have a nematic director
    oriented perpendicular to a weak bond orientation at $\theta \in \{ \pi/6,\pi/2,5\pi/6\}$
    and have $\langle \cos(6\theta)\rangle =-1$. 
    (d) The critical temperature as a function of Moir\'e lattice filling as determined by
    Monte Carlo. At $1/3-,2/5-,$ and $1/2-$ filling we find the same charge ordered states
    as in (b). Between $2/5-$filling and $1/2-$filling we find a type-I nematic state
    defined by short-range domains of the $1/2$-filled charge stripe state. Above
    $1/3$-filling, we find an isotropic state defined by hexagonal domains of the 
    $1/3$ generalized Wigner crystal, which eventually gives way to a type-II
    nematic state defined by fragmented domains of the $2/5$-filled columnar 
    dimer crystal.
    }
    \label{fig:setup}
\end{figure*}

    The promise of a highly tunable lattice system that can allow solid-state-based simulation of strong coupling physics~\cite{Andrei2021Nat.Rev.Mater.} 
    has largely driven the explosion of efforts studying Moir\'e superlattices.
    The transition metal dichalcogenide (TMD) heterobilayer Moir\'e systems
    with zero twist-angle (see Fig.~\ref{fig:setup}(a)) and localized Wannier orbitals
    form a uniquely simple platform to explore phases driven by strong interactions~\cite{Tang2020Nature,Regan2020Nature}. 
    Indeed, upon sweeping the density of electrons per Moir\'e unit cell, incompressible charge ordered states have 
    been observed at various fractional fillings~\cite{Regan2020Nature,Xu2020Nature,Li2021Naturea}. These charge orders can be viewed as a generalized Wigner crystalline state that reduces the symmetry of the underlying Moir\'e lattice, as they are driven by the long-range Coulomb interaction. The density controlled melting of Wigner crystals is expected to result in a rich hierarchy of intermediate phases~\cite{Spivak2004Phys.Rev.B,Spivak2006AnnalsofPhysics,Jamei2005Phys.Rev.Lett.}. While a microscopic theoretical study of Wigner crystal melting is challenging due to the continuous spatial symmetry, the melting of generalized Wigner crystals is more amenable to a microscopic study due to the lattice. The observation of intermediate compressible states with optical anisotropy~\cite{Jin2021Nat.Mater.} (see Fig.~\ref{fig:setup}(b)) and the tunability beyond density~\cite{Li2021Nature,Ghiotto2021Nature} present a tantalizing possibility to study melting and the possible intermediate phases of the generalized Wigner crystals.  
    
    The underlying lattice in the generalized Wigner crystal reduces the continuous rotational symmetry to $D_3$ point group symmetry.  
    Ref.~\cite{Coppersmith1982Phys.Rev.B} studied the melting of a $1/3$-filled crystalline state on a triangular lattice
    in the context of Krypton adsorbed on Graphene. 
    Based on the free energy costs of the domain walls and domain wall intersections, they reasoned that the generalized WC would first melt
    into a hexagonal liquid, and then crystallize into a stripe solid.
    From the modern perspectives of electronic
    liquid crystals~\cite{Kivelson1998Nature}, one anticipates nematic fluid states in the vicinity of crystalline anisotropic states such as
    the stripe solid. Moreover, the $D_3$ point group symmetry of the triangular
    lattice relevant for hetero-TMD Moir\'e sytems further enriches the possibilities of the intermediate fluid phases.
    The triangular lattice admits two types of nematic states
    due to the nematic order parameter transforming as a 2-dimensional irreducible representation of
    the lattice point group~\cite{Serre2012,Fernandes2020Sci.Adv.}.
    The hetero-TMD Moir\'e systems present an excellent opportunity to study these intermediate liquid phases.

    As the quantum melting of charge order is a notoriously difficult
    problem ~\cite{Kivelson2003Rev.Mod.Phys.},
    in this paper we will take advantage of the small bandwidth in hetero-TMD and use a strong coupling approach. 
    We study Monte Carlo simulations
    inspired by hetero-TMD Moir\'e systems at and between commensurate charge ordered states.
    We analyze our results in terms of the structure factor
    and a nematic order parameter correlation function.
    In particular, we distinguish between the two possible types of nematic states
    illustrated in Fig.~\ref{fig:setup}(c), one associated with the director aligned with a single majority bond orientation (which we dub type-I),
    and the other perpendicular to a single minority bond orientation (type-II).
    As shown in Fig.~\ref{fig:setup}(d), we
    find the type-I nematic 
    and type-II nematic to each robustly appear between 2/5 and 1/2 and 1/3 and 2/5 respectively. 
    We conclude with a discussion.

    As the orientation of the nematic director is defined within the angle range
    $\theta \in [0,\pi)$ (Fig. 1(c)) we define the local nematic field using
    complex notation $N(\vec{r}) = n(\vec{r}) e^{i 2\theta(\vec{r})}$, where 
    $n(\vec{r}) \in \mathbb{R}$. 
    In terms of this nematic order
    parameter field, the free energy density describing the isotropic-nematic transition in a trigonal system takes the following form
    \cite{Fernandes2020Sci.Adv.,Venderbos2018Phys.Rev.B,Hecker2018npjQuantMater,Little2020Nat.Mater.}: 
    \begin{align}
        \label{eqn:free_energy}
        f[N(\vec{r})] &= \frac{r}{2} \vert N(\vec{r}) \vert^4 + \frac{u}{4} \vert N(\vec{r}) \vert^2 + 
        \frac{\gamma}{6}(N(\vec{r})^3 + N^*(\vec{r})^3)\notag\\
        &=\frac{r}{2} n(\vec{r})^2 + \frac{u}{4} n(\vec{r})^4 + \frac{\gamma}{3} 
        n(\vec{r})^3\cos(6 \theta(\vec{r}))
    \end{align}
    The sign of the coefficient of the cubic term, $\gamma$, determines whether
    the system wants to be in a type-I nematic state ($\langle\cos(6\theta)\rangle = +1$) or type-II nematic
    state ($\langle\cos(6\theta)\rangle = -1$).

    We explore the phase diagram using classical Monte Carlo as a function of temperature $T$ and 
    the number of particles per Moir\'e site $\nu$. 
    To emulate the experimental setup in 
    refs.~\cite{Xu2020Nature,Jin2021Nat.Mater.,Li2021Nat.Mater.,Regan2020Nature}, 
    the Hamiltonian that we simulate describes the Coulomb interaction for
    electrons halfway between two dielectric gates a distance $d$ apart with dielectric constant $\epsilon$:
    \begin{align}
        \label{eqn:hamiltonian}
        \mathcal{H} = \frac{1}{2}\sum\limits_{i\neq j} \rho_i \rho_j
        \left(\frac{e^2}{4\pi \epsilon \epsilon_0 a}\right) \frac{4}{d} \sum\limits_{n = 0}^\infty 
        K_0 \left(\frac{\pi (2n+1)\vert\vec{r_i} - \vec{r_j}\vert}{d}\right).
    \end{align}
    Here, $K_0$ is the modified Bessel function of the second kind, $a$ is the Moir\'e 
    lattice constant, and $\rho_{i,j} \in \{0,1\}$ are the occupancies of lattice
    sites $i,j$. 
    As in refs.~\cite{Xu2020Nature,Jin2021Nat.Mater.}, 
    we take $a = 8nm$ and $d = 10a$, and     
    we take $e^2/(4\pi\epsilon\epsilon_0 a)$ as our unit of energy for simulation.

    Because the interaction is long-ranged, simply simulating a system with periodic
    boundary conditions would result in ambiguous distance calculations. 
    Thus, we simulate a formally infinite system
    that is constrained to be periodic in an $\ell \times \ell$ rhombus. Particles 
    interact both within and between copies of the system. 
    The choice of an $\ell \times \ell$ rhombus has the full symmetry of 
    the triangular lattice as, when one considers the infinite system,
    the action of each element of the point group is a bijective map on the set of
    unique sites contained within the simulation cell. Thus we do not expect our 
    choice of geometry to artificially promote rotational symmetry breaking.
    Moreover in each nematic state that we report,
    our simulations find configurations with each of the three possible director
    orientations for the relevant nematic type with equal probability.
    For Monte Carlo updates, we use arbitrary-range, single particle occupancy exchanges with standard Metropolis acceptance
    rules. However, the prevalence of short-range correlated structures leading to long autocorrelation times
    complicates our simulations, especially in the incompressible density region. To deal with this, we developed a cluster
    algorithm in the spirit of the well-known Wolff algorithm~\cite{Wolff1989Phys.Rev.Lett.}
    and the later geometric cluster algorithm~\cite{Heringa1998Phys.Rev.E}.
    For more detail, see appendix A. We perform a cluster update after every 1000 single particle occupancy exchange updates.

    At each point in phase space, we calculate the Monte Carlo average of the 
    structure factor
    \begin{align}
        \label{eqn:structure_factor}
        \langle S(\vec{Q}) \rangle =\frac{1}{\ell^2} \bigg\langle 
        \sum\limits_{i,j} \rho_i \rho_j e^{-i\vec{Q}\cdot(\vec{r}_i-\vec{r}_j)}
        \bigg\rangle,
    \end{align}
    to assess crystalline order.
    To assess the degree of rotational symmetry breaking, we also calculate the average
    of the nematic order parameter correlation function (per site) given by
    \begin{align}
        \label{eqn:corr_fn}
        \frac{1}{\ell^2}\sum\limits_{\vec{r},\vec{r'}}
        \langle C(\vec{r},\vec{r'}) \rangle = 
        \frac{1}{\ell^2}
        \sum\limits_{\vec{r},\vec{r'}} \langle N(\vec{r})N^*(\vec{r'})\rangle 
        = \frac{1}{\ell^2}\langle \tilde{C}(\vec{q} = 0) \rangle 
    \end{align}
    where $\vec{q}$ denotes Fourier momentum. 
    At high temperatures, when $\langle N(\vec{r}) \rangle = 0$,
    $\langle \tilde{C}(\vec{q} = 0) \rangle/\ell^2$ behaves as $k_B T$ times the nematic
    susceptibility: $\chi(\vec{q} = 0)k_B T$. Generically we expect this to have 
    some continuous behavior as a function of temperature.
    However, when the order parameter develops an expectation value in a nematic state,
    $\langle \tilde{C}(\vec{q} = 0) \rangle/\ell^2$ should acquire a constant,
    non-zero value.
    To determine the type of nematicity exhibited by nematic states, we also calculate
    $\langle \cos(6\theta) \rangle$, where, as in Fig.~\ref{fig:setup}(c), type-I (type-II) nematic states
    have $\langle \cos(6\theta) \rangle = +1\ (-1)$. For further details about
    the calculation of these quantities from our Monte Carlo simulation data, see
    appendix B. All results that we show are obtained from an $\ell = 20$ system,
    except for exactly at $\nu = 1/3$ since $20 \times 20/3$ is not an integer.
    In all cases, we perform $10^5$ updates per site for equilibration at each 
    temperature, and then $2\times 10^5$ updates per site for data collection.

\begin{figure}
    \centering
    \includegraphics[width=\linewidth]{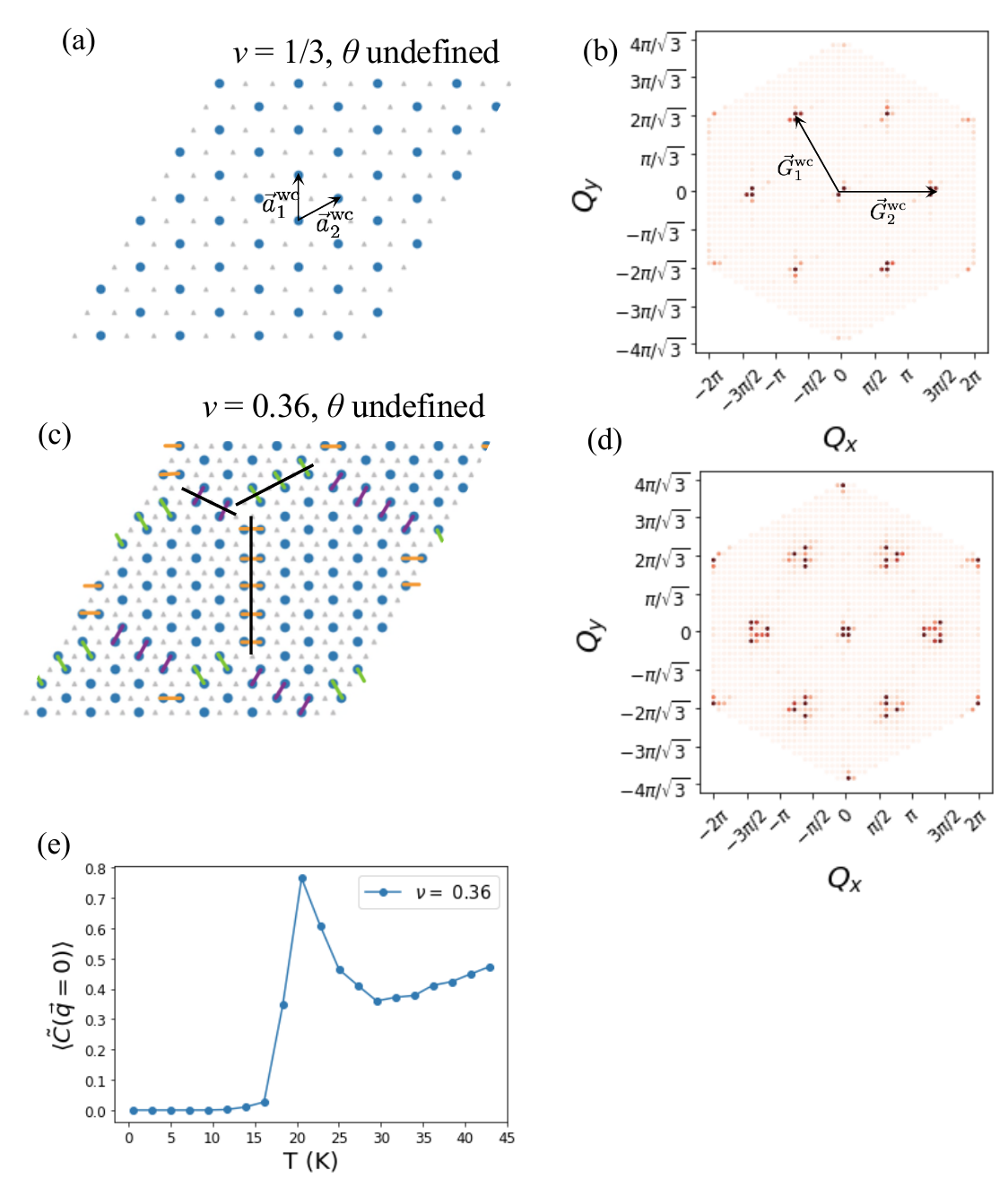}
    \caption{(a) Generalized Wigner crystal at $\nu=1/3$ particles per Moir\'e site for
    an $\ell = 12$ system obtained by Monte Carlo. 
    The system is isotropic and thus the orientation of the
    nematic director, $\theta$, is undefined. 
    (b) The Monte Carlo average of the structure factor at $\nu = 1/3$. 
    The structure factor exhibits
    peaks at the reciprocal lattice vectors of the $\nu = 1/3$ generalized 
    Wigner crystal.
    (c) Monte Carlo equilibrated state at $\nu = 0.36$ showing the isotropic hexagonal
    Wigner crystal domain state. Nearest neighbor bonds are shown and color-coded according
    to their orientation. Domain walls (marked with black lines) between the
    three registries of the generalized Wigner crystal state meet at $2\pi/3$ angles,
    forming hexagonal domains.
    (d) The Monte Carlo average of the structure factor at $\nu = 0.36$, showing the
    short-range correlated nature of the hexagonal Wigner crystal domain state in the
    broadened peaks as compared to (b).
    (e) The Monte Carlo average of the nematic correlation function at $\nu = 0.36$ 
    confirms the isotropic nature of the hexagonal Wigner crystal domain state, as it
    drops to zero below the phase transition. }
    \label{fig:isotropic_states}
\end{figure}

    At $\nu = 1/3$, we find the isotropic generalized Wigner crystalline phase, shown
    in Fig.~\ref{fig:isotropic_states}(a) for $\ell = 12$. This phase has lattice vectors
    $\vec{a}^{\text{wc}}_1 = (0,\sqrt{3})$ and $\vec{a}^{\text{wc}}_2 = (3/2,\sqrt{3}/2)$
    as indicated by the black arrows in Fig.~\ref{fig:isotropic_states}(a).
    The structure factor shows well defined peaks at the reciprocal lattice vectors 
    $\vec{G}^{\text{wc}}_1 = (-2\pi/3,2\pi/\sqrt{3}))$ and $\vec{G}^{\text{wc}}_2 = (4\pi/3,0)$
    associated with the  crystalline state (Fig.~\ref{fig:isotropic_states}(b)). 
    Upon increasing the density, this crystalline state starts to melt, 
    but it maintains an isotropic, compressible state to a certain filling. 
    At small fillings away from the $1/3$-state, as shown in 
    Fig.~\ref{fig:isotropic_states}(c), 
    the extra particles form domain walls between the three different registries of 
    the generalized WC state. Three domain walls are marked with black lines in 
    Fig.~\ref{fig:isotropic_states}(c).
    The domain walls meet at $2\pi/3$ angles, reminiscent
    of what was found in ref.~\cite{Coppersmith1982Phys.Rev.B}.

    As in ref.~\cite{Coppersmith1982Phys.Rev.B}, this domain wall 
    structure is stable while the density of domain walls is dilute (and hence
    the domain walls are long) because energetics favor $2\pi/3$ angles. 
    Taking into account interactions up to fifth neighbor, the energy contributed to
    the Hamiltonian by the six particles in the
    three dimers composing the $2\pi/3$ vertex is
    \begin{align}
        E_v = 3V_1 + \frac{21}{2} V_2 + 6 V_3 + \frac{21}{2} V_4 + \frac{15}{2} V_5
    \end{align}
    while that of the particles composing three straight domain wall dimers is
    \begin{align}
        E_{DW} = 3 V_1 + 12 V_2 + 6 V_3 + 6 V_4 + 9 V_5,
    \end{align}
    where $V_i$ denotes the energy of two $i$'th neighbor particles.
    Using eq.(~\ref{eqn:hamiltonian}) it is easy to check that
    $E_{DW} - E_v > 0$, and thus (at least to this order of interaction), 
    it is energetically favorable to have $2\pi/3$-vertices, even at low temperatures. 
    However, the densest possible hexagonal domain state
    consists of a close packing of $2\pi/3$-vertices, which has 
    density $\nu = 3/8$. Thus, this state certainly cannot exist at 
    densities $\nu > 3/8$.
    In Fig.~\ref{fig:isotropic_states}(d), 
    the structure factor of this compressible state is still dominated 
    by the generalized WC, but with the domain size
    becoming finite, the superstructure peaks are broadened. 
    Looking at the ensemble averaged nematic correlation function 
    $\langle \tilde{C}(\vec{q}=0) \rangle$ in Fig.~\ref{fig:isotropic_states}(e)
    confirms that this state is isotropic as it drops to zero below the transition.

\begin{figure*}
    \centering
    \includegraphics[width=\linewidth]{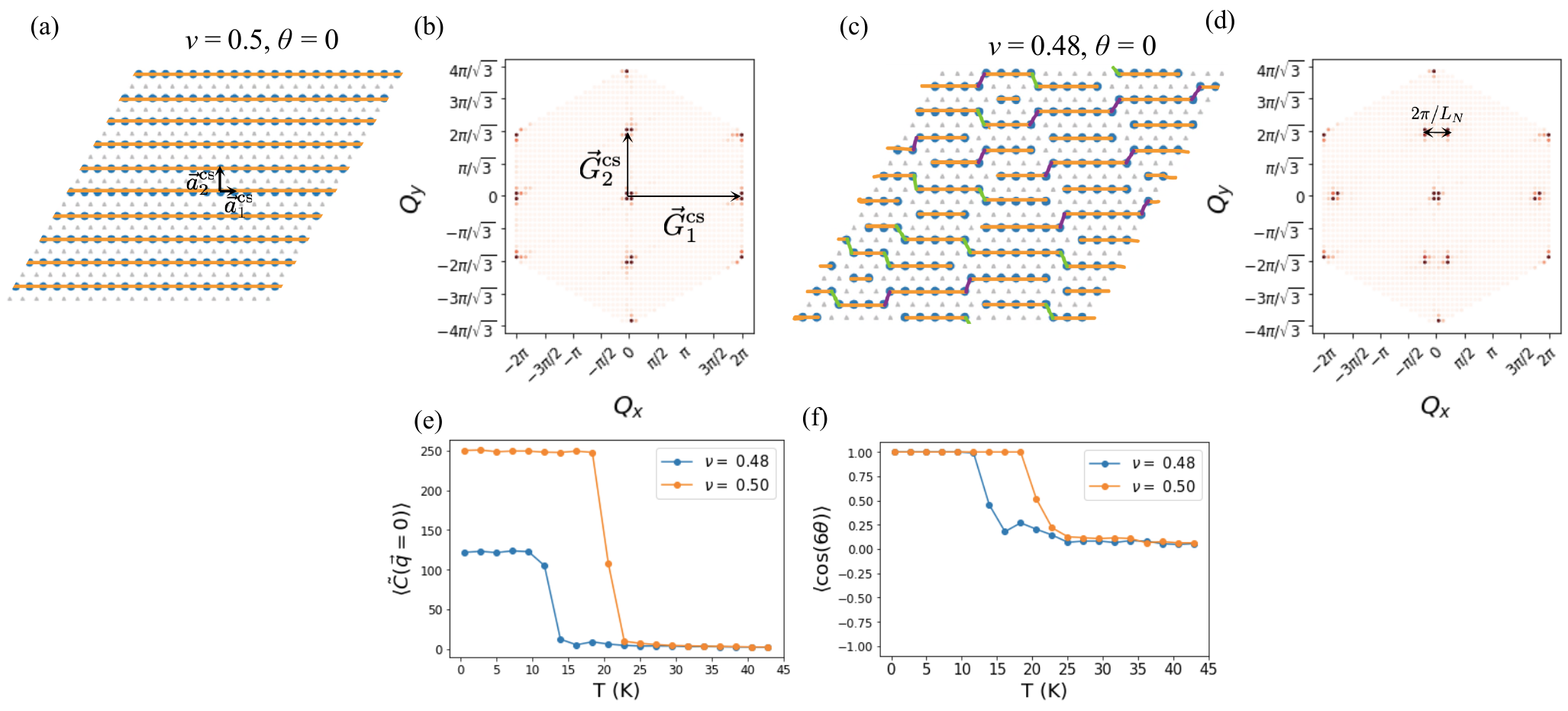}
    \caption{
        (a) The Monte Carlo charge stripe state at $\nu = 1/2$ 
        shown for nematic director orientation $\theta = 0$. 
        We annotate nearest neighbor bonds and color them according to their orientation.
        The red arrows indicate the charge stripe lattice vectors.
        (b) The Monte Carlo average of the structure factor at $\nu = 1/2$, averaged over 
        configurations with director orientation $\theta = 0$. 
        The red arrows indicate peaks at the reciprocal lattice vectors of the
        charge stripe state.
        (c) Monte Carlo equilibrated state at $\nu = 0.48$ showing short-ranged stripe
        nematic state for nematic director orientation $\theta = 0$. We again annotate
        nearest-neighbor bonds. 
        Pieces of the charge stripe state are separated by dislocations. 
        (d) The Monte Carlo average of the structure factor, at $\nu = 0.48$, averaged
        over configurations with director orientation $\theta = 0$.
        The peak at $(0,2\pi/\sqrt{3})$ splits into two peaks separated by $2\pi/L_N$ 
        where $L_N$ is the average stripe domain length.
        (e) The Monte Carlo average of the nematic order parameter correlation function 
        at $\nu = 1/2$ and $\nu = 0.48$. It jumps to a finite, constant value at $T_c$.
        (f) $\langle \cos(6\theta)\rangle$ at $\nu = 1/2$ and $\nu = 0.48$, which
        goes to $+1$ at $T_c$ in both cases. This suggests type-I nematicity at 
        both of these fillings.
    }
    \label{fig:typeI_nematic}
\end{figure*}

    The state is dramatically different at $\nu = 1/2$. We have the charge stripe
    state shown in Fig.~\ref{fig:typeI_nematic}(a) 
    with lattice vectors $\vec{a}^{\text{cs}}_1 = (1,0)$ and
    $\vec{a}^{\text{cs}}_2 = (0,\sqrt{3})$. 
    There are two degenerate charge stripe states whose lattice
    vectors are related by $\pi/3$ and $2\pi/3$ rotations of $\vec{a}^{\text{cs}}_{1,2}$.
    The structure factor in Fig.~\ref{fig:typeI_nematic}(b), 
    averaged over configurations with the same orientation as the one shown in 
    Fig.~\ref{fig:typeI_nematic}(a),
    contains peaks at the reciprocal lattice vectors 
    $\vec{G}^{\text{cs}}_1 = (2\pi,0)$ and
    $\vec{G}^{\text{cs}}_2 = (0,2\pi/\sqrt{3})$. As expected,
    $\vec{G}^{\text{cs}}_i \cdot \vec{a}^{\text{cs}}_j = 2\pi \delta_{ij}$.
    Diluting the $1/2$-filled state,
    the stripes become shorter via the introduction of dislocations, as shown
    in Fig~\ref{fig:typeI_nematic}(c). 
    The structure factor reflects the finite length of these stripe domains in the 
    splitting of the stripe peak over the span of the stripe domain size scale. 
    The peak at $\vec{a}^{\text{cs}}_2$ is split into two peaks separated by $2\pi / L_N$
    where $L_N \approx 4.63$ is the average stripe domain size.
    The nematic correlation function reveals that, unlike the isotropic phases
    in Fig.~\ref{fig:isotropic_states}(e), 
    $\langle \tilde{C}(\vec{q} = 0)\rangle$ shows a sharp jump at $T_c$ 
    to a finite value in Fig.~\ref{fig:typeI_nematic}(e). 
    This indicates that these are nematic states.
    By examining $\langle \cos(6\theta) \rangle$ in Fig.~\ref{fig:typeI_nematic}(f), 
    we can see that both of these phases are type-I nematic.

\begin{figure*}
    \centering
    \includegraphics[width=\linewidth]{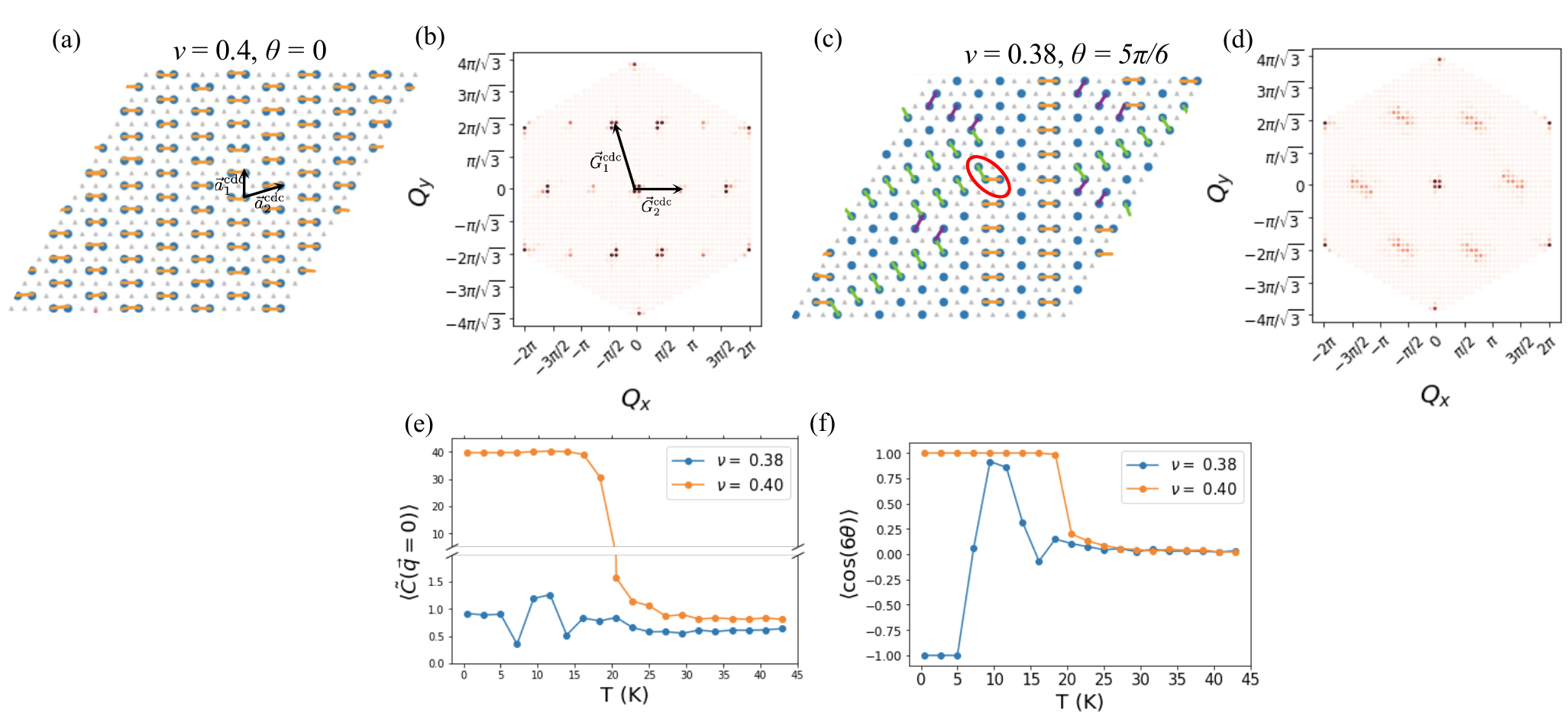}
    \caption{
    (a) The columnar dimer crystal state obtained from Monte Carlo simulations at
    $\nu = 0.4$, shown with nematic director orientation $\theta = 0$. We annotate
    the nearest neighbor bonds and color them according to orientation.
    The red arrows indicate the columnar dimer crystal lattice vectors.
    (b) The Monte Carlo average of the structure factor at $\nu = 2/5$, averaged over 
    configurations with director orientation $\theta = 0$. The red arrows indicate 
    peaks at the reciprocal lattice vectors of the columnar dimer crystal state.
    (c) The fragmented dimer column state at $\nu = 0.38$, shown with nematic director
    orientation $\theta = \pi / 6$.
    (d) The Monte Carlo average of the structure factor at $\nu = 0.38$, averaged
    over configurations with director orientation $\theta = \pi/6$. Broad peaks
    at the reciprocal lattice vectors of the generalized Wigner crystal state 
    appear due to the short-range correlated regions of generalized Wigner crystal
    between the dimer column fragments.
    (e) The nematic order parameter correlation function is finite and constant at
    low temperatures, showing that these are nematic states. 
    (f) $\langle \cos(6\theta) \rangle$ for $\nu = 2/5$ and $\nu = 0.38$.
    The columnar dimer crystal has type-I nematicity as 
    $\langle \cos(6\theta) \rangle = +1$ at low-T. The fragmented
    dimer column state is a type-II nematic as
    $\langle \cos(6\theta) \rangle = +1$ at low-T.
    }
    \label{fig:typeII_nematic}
\end{figure*}

    Upon further diluting, the system maintains the same type of anisotropy and forms the columnar dimer crystal state at $\nu=2/5$,
    shown in Fig.~\ref{fig:typeII_nematic}(a) with director orientation $\theta = 0$.
    This is the limit of the shortest stripe length, evolving from $\nu=1/2$. 
    The $\nu = 2/5$ state is a crystalline state 
    with lattice vectors $\vec{a}^\text{cdc}_1 = (0,\sqrt{3})$ and 
    $\vec{a}^\text{cdc}_2 = (5/2,\sqrt{3}/2)$.
    We mark the reciprocal lattice vectors $\vec{G}^\text{cdc}_1 = (-2\pi/5,2\pi/\sqrt{3})$ and
    $\vec{G}^\text{cdc}_2 = (4\pi/5,0)$
    in the structure factor shown in Fig.~\ref{fig:typeII_nematic}(b). 
    Note that the peak at $2\vec{G}^\text{cdc}_2$ is more intense than the one at
    $\vec{G}^\text{cdc}_2$. This is due to the form factor from the lattice basis.
    As we dilute further, the length of the
    columns get shorter as dimers get broken up. At lower densities, the
    columns do not extend over the entire system, so there are finite length
    segments of columns that can have different orientations as illustrated in Fig.~\ref{fig:typeII_nematic}(c).
    The broken pieces of dimers form short-range correlated domains of generalized WC.
    This is shown by the broad peaks in the structure factor in 
    Fig.~\ref{fig:typeII_nematic}(d). This compressible state no longer has the mirror symmetries
    of the columnar state. It is still anisotropic as we can see from the
    nematic correlation function in Fig.~\ref{fig:typeII_nematic}(e).
    Interestingly, the columnar fragments intersect at $2\pi/3$ angles, as well as $\pi/3$ angles,
    one of which is circled in red in fig.~\ref{fig:typeII_nematic}(c).
    While the $2\pi/3$ intersections are isotropic,
    the $\pi/3$ intersections consist 
    primarily of only two of the three possible nearest-neighbor bond orientations, and hence this state is
    a type-II nematic phase. We confirm this by observing that 
    $\langle \cos(6\theta) \rangle = -1$ at low temperatures in 
    Fig.~\ref{fig:typeII_nematic}(f). Thus we predict microscopic mechanism for the type-II nematic phase.

    The nematic-II state in the region of $3/8 < \nu < 2/5$ is supported energetically.
    Upon increasing density beyond $\nu > 3/8$, columnar fragments have to either intersect
    also at $\pi/3$ or be parallel to each other.
    Since the distance between columnar fragments increases away from the $\pi/3$ intersection,
    we expect $\pi/3$ intersections to be favored. See appendix C for a schematic calculation
    demonstrating this.
    Such $\pi/3$ intersections involve two nearest-neighbor bond orientations, promoting
    a nematic-II state. 

    One could experimentally probe our predicted nematic states by performing optical measurements similar
    to those done at $\nu = 1/2$ in ref.~\cite{Jin2021Nat.Mater.}. 
    As one lowers the density from $\nu = 1/2$ to $\nu = 1/3$ we would anticipate
    a rotation of the nematic director and consequently a shift in the peaks of the 
    measured optical anisotropy axis. In particular, as one decreases the density
    from between $\nu = 1/2$ and $\nu = 2/5$, we predict that the measured
    anisotropy axis should have peaks along the nematic-I orientations
    $0, \pi/3, 2\pi/3$. Below $\nu =2/5$, when the director rotates into the
    nematic-II state, the peaks should be at $\pi/6, \pi/2, 5\pi/6$.
    Finally, below $\nu = 3/8$ when the system becomes isotropic, we expect
    that there should be no preferred anisotropy axis at all.
    For $1/3 < \nu < 3/8$, one could also look for signatures of the hexagonal WC domain state
    using Umklapp spectroscopy experiments like those done in ref.~\cite{Shimazaki2021Phys.Rev.X}. The short-range
    correlated nature of this state should show up as broadened Umklapp resonances
    around the  $\nu=1/3$ generalized Wigner crystal lattice vectors.

    In summary, we studied the electronic states of a system of strongly correlated electrons
    on a triangular lattice in the region $1/3 \leq \nu \leq 1/2$ particles per Moir\'e site.
    At $\nu = 1/2$, we find the charge stripe state. Upon dilution, the charge stripe
    state melts into a nematic-I short-ranged charge stripe state via the introduction of dislocations. Once the stripes become short enough, the columnar dimer crystal state emerges at $\nu = 2/5$. At even lower densities, the remaining columnar fragments space
    themselves out to lower their energy by intersecting at $\pi/3$ and $2\pi/3$ angles, resulting in a nematic-II state. 
    Below $\nu = 3/8$, the system can again lower its energy by using only $2\pi/3$ columnar fragment intersections
    to form an isotropic,
    hexagonal network of domain walls between regions of the $\nu = 1/3$ generalized Wigner crystal. Finally, at $\nu = 1/3$, the pure, isotropic generalized Wigner crystal state emerges.
    Our intermediate states were not only promoted by entropy, but we also found them to have lower energy
    compared to macroscopically phase separated states.
    Accordingly, we suspect that our proposed states are relevant for finite experimental
    temperatures where fluctuations due to entropy also play a role. We leave the determination of the classical ground state at $T=0$ 
    as a subject of future work.

    Studying the intermediate phases of melted density waves has been of interest since considerations of Krypton adsorbed on graphene~\cite{Coppersmith1982Phys.Rev.B}. 
    However, limitations in computational resources and experimental methods caused
    difficulties in probing the intermediate states.
    With advances in computing power and the advent of the TMD Moir\'e platform, however, detailed phase diagrams can now be predicted computationally
    and probed experimentally. Our work demonstrates this capacity to explore
    intermediate phases and the richness of the phase diagram one can obtain with
    a classical model, even without considering quantum effects. 
    We found the striped phase predicted upon increasing density in ref.~\cite{Coppersmith1982Phys.Rev.B} refines into
    two distinct stripe crystal states neighboring two distinct types of nematics. In particular, we presented 
    a microscopic mechanism for the formation of the nematic-II state
    via $\pi/3$ intersections between columnar fragments. As a subject of future
    work, it would be interesting to study the implications of our findings for the
    melting of WCs without a lattice potential, such as those recently observed
    in~\cite{Zhou2021Nature,Smolenski2021Nature}.
    
{\bf Acknowledgements:} The authors acknowledge support by the NSF [Platform for the Accelerated Realization, Analysis, and Discovery of Interface Materials (PARADIM)] under cooperative agreement no. DMR-U638986.
We would also like to thank Steven Kivelson,
Kin-Fai Mak, Jie Shan, Sue Coppersmith, Ata\c{c} Imamo\u{g}lu, and Samuel Lederer for helpful discussions.

\bibliography{tmd_heterobilayer_citations}

\begin{thebibliography}{25}%
\makeatletter
\providecommand \@ifxundefined [1]{%
 \@ifx{#1\undefined}
}%
\providecommand \@ifnum [1]{%
 \ifnum #1\expandafter \@firstoftwo
 \else \expandafter \@secondoftwo
 \fi
}%
\providecommand \@ifx [1]{%
 \ifx #1\expandafter \@firstoftwo
 \else \expandafter \@secondoftwo
 \fi
}%
\providecommand \natexlab [1]{#1}%
\providecommand \enquote  [1]{``#1''}%
\providecommand \bibnamefont  [1]{#1}%
\providecommand \bibfnamefont [1]{#1}%
\providecommand \citenamefont [1]{#1}%
\providecommand \href@noop [0]{\@secondoftwo}%
\providecommand \href [0]{\begingroup \@sanitize@url \@href}%
\providecommand \@href[1]{\@@startlink{#1}\@@href}%
\providecommand \@@href[1]{\endgroup#1\@@endlink}%
\providecommand \@sanitize@url [0]{\catcode `\\12\catcode `\$12\catcode
  `\&12\catcode `\#12\catcode `\^12\catcode `\_12\catcode `\%12\relax}%
\providecommand \@@startlink[1]{}%
\providecommand \@@endlink[0]{}%
\providecommand \url  [0]{\begingroup\@sanitize@url \@url }%
\providecommand \@url [1]{\endgroup\@href {#1}{\urlprefix }}%
\providecommand \urlprefix  [0]{URL }%
\providecommand \Eprint [0]{\href }%
\providecommand \doibase [0]{http://dx.doi.org/}%
\providecommand \selectlanguage [0]{\@gobble}%
\providecommand \bibinfo  [0]{\@secondoftwo}%
\providecommand \bibfield  [0]{\@secondoftwo}%
\providecommand \translation [1]{[#1]}%
\providecommand \BibitemOpen [0]{}%
\providecommand \bibitemStop [0]{}%
\providecommand \bibitemNoStop [0]{.\EOS\space}%
\providecommand \EOS [0]{\spacefactor3000\relax}%
\providecommand \BibitemShut  [1]{\csname bibitem#1\endcsname}%
\let\auto@bib@innerbib\@empty
\bibitem [{\citenamefont {Andrei}\ \emph {et~al.}(2021)\citenamefont {Andrei},
  \citenamefont {Efetov}, \citenamefont {{Jarillo-Herrero}}, \citenamefont
  {MacDonald}, \citenamefont {Mak}, \citenamefont {Senthil}, \citenamefont
  {Tutuc}, \citenamefont {Yazdani},\ and\ \citenamefont
  {Young}}]{Andrei2021Nat.Rev.Mater.}%
  \BibitemOpen
  \bibfield  {author} {\bibinfo {author} {\bibfnamefont {Eva~Y.}\ \bibnamefont
  {Andrei}}, \bibinfo {author} {\bibfnamefont {Dmitri~K.}\ \bibnamefont
  {Efetov}}, \bibinfo {author} {\bibfnamefont {Pablo}\ \bibnamefont
  {{Jarillo-Herrero}}}, \bibinfo {author} {\bibfnamefont {Allan~H.}\
  \bibnamefont {MacDonald}}, \bibinfo {author} {\bibfnamefont {Kin~Fai}\
  \bibnamefont {Mak}}, \bibinfo {author} {\bibfnamefont {T.}~\bibnamefont
  {Senthil}}, \bibinfo {author} {\bibfnamefont {Emanuel}\ \bibnamefont
  {Tutuc}}, \bibinfo {author} {\bibfnamefont {Ali}\ \bibnamefont {Yazdani}}, \
  and\ \bibinfo {author} {\bibfnamefont {Andrea~F.}\ \bibnamefont {Young}},\
  }\bibfield  {title} {\enquote {\bibinfo {title} {The marvels of moir\'e
  materials},}\ }\href {\doibase 10.1038/s41578-021-00284-1} {\bibfield
  {journal} {\bibinfo  {journal} {Nature Reviews Materials}\ }\textbf {\bibinfo
  {volume} {6}},\ \bibinfo {pages} {201--206} (\bibinfo {year}
  {2021})}\BibitemShut {NoStop}%
\bibitem [{\citenamefont {Tang}\ \emph {et~al.}(2020)\citenamefont {Tang},
  \citenamefont {Li}, \citenamefont {Li}, \citenamefont {Xu}, \citenamefont
  {Liu}, \citenamefont {Barmak}, \citenamefont {Watanabe}, \citenamefont
  {Taniguchi}, \citenamefont {MacDonald}, \citenamefont {Shan},\ and\
  \citenamefont {Mak}}]{Tang2020Nature}%
  \BibitemOpen
  \bibfield  {author} {\bibinfo {author} {\bibfnamefont {Yanhao}\ \bibnamefont
  {Tang}}, \bibinfo {author} {\bibfnamefont {Lizhong}\ \bibnamefont {Li}},
  \bibinfo {author} {\bibfnamefont {Tingxin}\ \bibnamefont {Li}}, \bibinfo
  {author} {\bibfnamefont {Yang}\ \bibnamefont {Xu}}, \bibinfo {author}
  {\bibfnamefont {Song}\ \bibnamefont {Liu}}, \bibinfo {author} {\bibfnamefont
  {Katayun}\ \bibnamefont {Barmak}}, \bibinfo {author} {\bibfnamefont {Kenji}\
  \bibnamefont {Watanabe}}, \bibinfo {author} {\bibfnamefont {Takashi}\
  \bibnamefont {Taniguchi}}, \bibinfo {author} {\bibfnamefont {Allan~H.}\
  \bibnamefont {MacDonald}}, \bibinfo {author} {\bibfnamefont {Jie}\
  \bibnamefont {Shan}}, \ and\ \bibinfo {author} {\bibfnamefont {Kin~Fai}\
  \bibnamefont {Mak}},\ }\bibfield  {title} {\enquote {\bibinfo {title}
  {Simulation of {{Hubbard}} model physics in {{WSe2}}/{{WS2}}
  moir\'esuperlattices},}\ }\href {\doibase 10.1038/s41586-020-2085-3}
  {\bibfield  {journal} {\bibinfo  {journal} {Nature}\ }\textbf {\bibinfo
  {volume} {579}},\ \bibinfo {pages} {353--358} (\bibinfo {year}
  {2020})}\BibitemShut {NoStop}%
\bibitem [{\citenamefont {Regan}\ \emph {et~al.}(2020)\citenamefont {Regan},
  \citenamefont {Wang}, \citenamefont {Jin}, \citenamefont {Bakti~Utama},
  \citenamefont {Gao}, \citenamefont {Wei}, \citenamefont {Zhao}, \citenamefont
  {Zhao}, \citenamefont {Zhang}, \citenamefont {Yumigeta}, \citenamefont
  {Blei}, \citenamefont {Carlstr{\"o}m}, \citenamefont {Watanabe},
  \citenamefont {Taniguchi}, \citenamefont {Tongay}, \citenamefont {Crommie},
  \citenamefont {Zettl},\ and\ \citenamefont {Wang}}]{Regan2020Nature}%
  \BibitemOpen
  \bibfield  {author} {\bibinfo {author} {\bibfnamefont {Emma~C.}\ \bibnamefont
  {Regan}}, \bibinfo {author} {\bibfnamefont {Danqing}\ \bibnamefont {Wang}},
  \bibinfo {author} {\bibfnamefont {Chenhao}\ \bibnamefont {Jin}}, \bibinfo
  {author} {\bibfnamefont {M.~Iqbal}\ \bibnamefont {Bakti~Utama}}, \bibinfo
  {author} {\bibfnamefont {Beini}\ \bibnamefont {Gao}}, \bibinfo {author}
  {\bibfnamefont {Xin}\ \bibnamefont {Wei}}, \bibinfo {author} {\bibfnamefont
  {Sihan}\ \bibnamefont {Zhao}}, \bibinfo {author} {\bibfnamefont {Wenyu}\
  \bibnamefont {Zhao}}, \bibinfo {author} {\bibfnamefont {Zuocheng}\
  \bibnamefont {Zhang}}, \bibinfo {author} {\bibfnamefont {Kentaro}\
  \bibnamefont {Yumigeta}}, \bibinfo {author} {\bibfnamefont {Mark}\
  \bibnamefont {Blei}}, \bibinfo {author} {\bibfnamefont {Johan~D.}\
  \bibnamefont {Carlstr{\"o}m}}, \bibinfo {author} {\bibfnamefont {Kenji}\
  \bibnamefont {Watanabe}}, \bibinfo {author} {\bibfnamefont {Takashi}\
  \bibnamefont {Taniguchi}}, \bibinfo {author} {\bibfnamefont {Sefaattin}\
  \bibnamefont {Tongay}}, \bibinfo {author} {\bibfnamefont {Michael}\
  \bibnamefont {Crommie}}, \bibinfo {author} {\bibfnamefont {Alex}\
  \bibnamefont {Zettl}}, \ and\ \bibinfo {author} {\bibfnamefont {Feng}\
  \bibnamefont {Wang}},\ }\bibfield  {title} {\enquote {\bibinfo {title} {Mott
  and generalized {{Wigner}} crystal states in {{WSe2}}/{{WS2}}
  moir\'esuperlattices},}\ }\href {\doibase 10.1038/s41586-020-2092-4}
  {\bibfield  {journal} {\bibinfo  {journal} {Nature}\ }\textbf {\bibinfo
  {volume} {579}},\ \bibinfo {pages} {359--363} (\bibinfo {year}
  {2020})}\BibitemShut {NoStop}%
\bibitem [{\citenamefont {Xu}\ \emph {et~al.}(2020)\citenamefont {Xu},
  \citenamefont {Liu}, \citenamefont {Rhodes}, \citenamefont {Watanabe},
  \citenamefont {Taniguchi}, \citenamefont {Hone}, \citenamefont {Elser},
  \citenamefont {Mak},\ and\ \citenamefont {Shan}}]{Xu2020Nature}%
  \BibitemOpen
  \bibfield  {author} {\bibinfo {author} {\bibfnamefont {Yang}\ \bibnamefont
  {Xu}}, \bibinfo {author} {\bibfnamefont {Song}\ \bibnamefont {Liu}}, \bibinfo
  {author} {\bibfnamefont {Daniel~A.}\ \bibnamefont {Rhodes}}, \bibinfo
  {author} {\bibfnamefont {Kenji}\ \bibnamefont {Watanabe}}, \bibinfo {author}
  {\bibfnamefont {Takashi}\ \bibnamefont {Taniguchi}}, \bibinfo {author}
  {\bibfnamefont {James}\ \bibnamefont {Hone}}, \bibinfo {author}
  {\bibfnamefont {Veit}\ \bibnamefont {Elser}}, \bibinfo {author}
  {\bibfnamefont {Kin~Fai}\ \bibnamefont {Mak}}, \ and\ \bibinfo {author}
  {\bibfnamefont {Jie}\ \bibnamefont {Shan}},\ }\bibfield  {title} {\enquote
  {\bibinfo {title} {Correlated insulating states at fractional fillings of
  moir\'esuperlattices},}\ }\href {\doibase 10.1038/s41586-020-2868-6}
  {\bibfield  {journal} {\bibinfo  {journal} {Nature}\ }\textbf {\bibinfo
  {volume} {587}},\ \bibinfo {pages} {214--218} (\bibinfo {year}
  {2020})}\BibitemShut {NoStop}%
\bibitem [{\citenamefont {Li}\ \emph {et~al.}(2021{\natexlab{a}})\citenamefont
  {Li}, \citenamefont {Li}, \citenamefont {Regan}, \citenamefont {Wang},
  \citenamefont {Zhao}, \citenamefont {Kahn}, \citenamefont {Yumigeta},
  \citenamefont {Blei}, \citenamefont {Taniguchi}, \citenamefont {Watanabe},
  \citenamefont {Tongay}, \citenamefont {Zettl}, \citenamefont {Crommie},\ and\
  \citenamefont {Wang}}]{Li2021Naturea}%
  \BibitemOpen
  \bibfield  {author} {\bibinfo {author} {\bibfnamefont {Hongyuan}\
  \bibnamefont {Li}}, \bibinfo {author} {\bibfnamefont {Shaowei}\ \bibnamefont
  {Li}}, \bibinfo {author} {\bibfnamefont {Emma~C.}\ \bibnamefont {Regan}},
  \bibinfo {author} {\bibfnamefont {Danqing}\ \bibnamefont {Wang}}, \bibinfo
  {author} {\bibfnamefont {Wenyu}\ \bibnamefont {Zhao}}, \bibinfo {author}
  {\bibfnamefont {Salman}\ \bibnamefont {Kahn}}, \bibinfo {author}
  {\bibfnamefont {Kentaro}\ \bibnamefont {Yumigeta}}, \bibinfo {author}
  {\bibfnamefont {Mark}\ \bibnamefont {Blei}}, \bibinfo {author} {\bibfnamefont
  {Takashi}\ \bibnamefont {Taniguchi}}, \bibinfo {author} {\bibfnamefont
  {Kenji}\ \bibnamefont {Watanabe}}, \bibinfo {author} {\bibfnamefont
  {Sefaattin}\ \bibnamefont {Tongay}}, \bibinfo {author} {\bibfnamefont {Alex}\
  \bibnamefont {Zettl}}, \bibinfo {author} {\bibfnamefont {Michael~F.}\
  \bibnamefont {Crommie}}, \ and\ \bibinfo {author} {\bibfnamefont {Feng}\
  \bibnamefont {Wang}},\ }\bibfield  {title} {\enquote {\bibinfo {title}
  {Imaging two-dimensional generalized {{Wigner}} crystals},}\ }\href {\doibase
  10.1038/s41586-021-03874-9} {\bibfield  {journal} {\bibinfo  {journal}
  {Nature}\ }\textbf {\bibinfo {volume} {597}},\ \bibinfo {pages} {650--654}
  (\bibinfo {year} {2021}{\natexlab{a}})}\BibitemShut {NoStop}%
\bibitem [{\citenamefont {Spivak}\ and\ \citenamefont
  {Kivelson}(2004)}]{Spivak2004Phys.Rev.B}%
  \BibitemOpen
  \bibfield  {author} {\bibinfo {author} {\bibfnamefont {Boris}\ \bibnamefont
  {Spivak}}\ and\ \bibinfo {author} {\bibfnamefont {Steven~A.}\ \bibnamefont
  {Kivelson}},\ }\bibfield  {title} {\enquote {\bibinfo {title} {Phases
  intermediate between a two-dimensional electron liquid and {{Wigner}}
  crystal},}\ }\href {\doibase 10.1103/PhysRevB.70.155114} {\bibfield
  {journal} {\bibinfo  {journal} {Phys. Rev. B}\ }\textbf {\bibinfo {volume}
  {70}},\ \bibinfo {pages} {155114} (\bibinfo {year} {2004})}\BibitemShut
  {NoStop}%
\bibitem [{\citenamefont {Spivak}\ and\ \citenamefont
  {Kivelson}(2006)}]{Spivak2006AnnalsofPhysics}%
  \BibitemOpen
  \bibfield  {author} {\bibinfo {author} {\bibfnamefont {Boris}\ \bibnamefont
  {Spivak}}\ and\ \bibinfo {author} {\bibfnamefont {Steven~A.}\ \bibnamefont
  {Kivelson}},\ }\bibfield  {title} {\enquote {\bibinfo {title} {Transport in
  two dimensional electronic micro-emulsions},}\ }\href {\doibase
  10.1016/j.aop.2005.12.002} {\bibfield  {journal} {\bibinfo  {journal} {Annals
  of Physics}\ }\textbf {\bibinfo {volume} {321}},\ \bibinfo {pages}
  {2071--2115} (\bibinfo {year} {2006})}\BibitemShut {NoStop}%
\bibitem [{\citenamefont {Jamei}\ \emph {et~al.}(2005)\citenamefont {Jamei},
  \citenamefont {Kivelson},\ and\ \citenamefont
  {Spivak}}]{Jamei2005Phys.Rev.Lett.}%
  \BibitemOpen
  \bibfield  {author} {\bibinfo {author} {\bibfnamefont {Reza}\ \bibnamefont
  {Jamei}}, \bibinfo {author} {\bibfnamefont {Steven}\ \bibnamefont
  {Kivelson}}, \ and\ \bibinfo {author} {\bibfnamefont {Boris}\ \bibnamefont
  {Spivak}},\ }\bibfield  {title} {\enquote {\bibinfo {title} {Universal
  {{Aspects}} of {{Coulomb-Frustrated Phase Separation}}},}\ }\href {\doibase
  10.1103/PhysRevLett.94.056805} {\bibfield  {journal} {\bibinfo  {journal}
  {Phys. Rev. Lett.}\ }\textbf {\bibinfo {volume} {94}},\ \bibinfo {pages}
  {056805} (\bibinfo {year} {2005})}\BibitemShut {NoStop}%
\bibitem [{\citenamefont {Jin}\ \emph {et~al.}(2021)\citenamefont {Jin},
  \citenamefont {Tao}, \citenamefont {Li}, \citenamefont {Xu}, \citenamefont
  {Tang}, \citenamefont {Zhu}, \citenamefont {Liu}, \citenamefont {Watanabe},
  \citenamefont {Taniguchi}, \citenamefont {Hone}, \citenamefont {Fu},
  \citenamefont {Shan},\ and\ \citenamefont {Mak}}]{Jin2021Nat.Mater.}%
  \BibitemOpen
  \bibfield  {author} {\bibinfo {author} {\bibfnamefont {Chenhao}\ \bibnamefont
  {Jin}}, \bibinfo {author} {\bibfnamefont {Zui}\ \bibnamefont {Tao}}, \bibinfo
  {author} {\bibfnamefont {Tingxin}\ \bibnamefont {Li}}, \bibinfo {author}
  {\bibfnamefont {Yang}\ \bibnamefont {Xu}}, \bibinfo {author} {\bibfnamefont
  {Yanhao}\ \bibnamefont {Tang}}, \bibinfo {author} {\bibfnamefont {Jiacheng}\
  \bibnamefont {Zhu}}, \bibinfo {author} {\bibfnamefont {Song}\ \bibnamefont
  {Liu}}, \bibinfo {author} {\bibfnamefont {Kenji}\ \bibnamefont {Watanabe}},
  \bibinfo {author} {\bibfnamefont {Takashi}\ \bibnamefont {Taniguchi}},
  \bibinfo {author} {\bibfnamefont {James~C.}\ \bibnamefont {Hone}}, \bibinfo
  {author} {\bibfnamefont {Liang}\ \bibnamefont {Fu}}, \bibinfo {author}
  {\bibfnamefont {Jie}\ \bibnamefont {Shan}}, \ and\ \bibinfo {author}
  {\bibfnamefont {Kin~Fai}\ \bibnamefont {Mak}},\ }\bibfield  {title} {\enquote
  {\bibinfo {title} {Stripe phases in {{WSe2}}/{{WS2}} moir\'e
  superlattices},}\ }\href {\doibase 10.1038/s41563-021-00959-8} {\bibfield
  {journal} {\bibinfo  {journal} {Nat. Mater.}\ }\textbf {\bibinfo {volume}
  {20}},\ \bibinfo {pages} {940--944} (\bibinfo {year} {2021})}\BibitemShut
  {NoStop}%
\bibitem [{\citenamefont {Li}\ \emph {et~al.}(2021{\natexlab{b}})\citenamefont
  {Li}, \citenamefont {Jiang}, \citenamefont {Li}, \citenamefont {Zhang},
  \citenamefont {Kang}, \citenamefont {Zhu}, \citenamefont {Watanabe},
  \citenamefont {Taniguchi}, \citenamefont {Chowdhury}, \citenamefont {Fu},
  \citenamefont {Shan},\ and\ \citenamefont {Mak}}]{Li2021Nature}%
  \BibitemOpen
  \bibfield  {author} {\bibinfo {author} {\bibfnamefont {Tingxin}\ \bibnamefont
  {Li}}, \bibinfo {author} {\bibfnamefont {Shengwei}\ \bibnamefont {Jiang}},
  \bibinfo {author} {\bibfnamefont {Lizhong}\ \bibnamefont {Li}}, \bibinfo
  {author} {\bibfnamefont {Yang}\ \bibnamefont {Zhang}}, \bibinfo {author}
  {\bibfnamefont {Kaifei}\ \bibnamefont {Kang}}, \bibinfo {author}
  {\bibfnamefont {Jiacheng}\ \bibnamefont {Zhu}}, \bibinfo {author}
  {\bibfnamefont {Kenji}\ \bibnamefont {Watanabe}}, \bibinfo {author}
  {\bibfnamefont {Takashi}\ \bibnamefont {Taniguchi}}, \bibinfo {author}
  {\bibfnamefont {Debanjan}\ \bibnamefont {Chowdhury}}, \bibinfo {author}
  {\bibfnamefont {Liang}\ \bibnamefont {Fu}}, \bibinfo {author} {\bibfnamefont
  {Jie}\ \bibnamefont {Shan}}, \ and\ \bibinfo {author} {\bibfnamefont
  {Kin~Fai}\ \bibnamefont {Mak}},\ }\bibfield  {title} {\enquote {\bibinfo
  {title} {Continuous {{Mott}} transition in semiconductor moir\'e
  superlattices},}\ }\href {\doibase 10.1038/s41586-021-03853-0} {\bibfield
  {journal} {\bibinfo  {journal} {Nature}\ }\textbf {\bibinfo {volume} {597}},\
  \bibinfo {pages} {350--354} (\bibinfo {year}
  {2021}{\natexlab{b}})}\BibitemShut {NoStop}%
\bibitem [{\citenamefont {Ghiotto}\ \emph {et~al.}(2021)\citenamefont
  {Ghiotto}, \citenamefont {Shih}, \citenamefont {Pereira}, \citenamefont
  {Rhodes}, \citenamefont {Kim}, \citenamefont {Zang}, \citenamefont {Millis},
  \citenamefont {Watanabe}, \citenamefont {Taniguchi}, \citenamefont {Hone},
  \citenamefont {Wang}, \citenamefont {Dean},\ and\ \citenamefont
  {Pasupathy}}]{Ghiotto2021Nature}%
  \BibitemOpen
  \bibfield  {author} {\bibinfo {author} {\bibfnamefont {Augusto}\ \bibnamefont
  {Ghiotto}}, \bibinfo {author} {\bibfnamefont {En-Min}\ \bibnamefont {Shih}},
  \bibinfo {author} {\bibfnamefont {Giancarlo S. S.~G.}\ \bibnamefont
  {Pereira}}, \bibinfo {author} {\bibfnamefont {Daniel~A.}\ \bibnamefont
  {Rhodes}}, \bibinfo {author} {\bibfnamefont {Bumho}\ \bibnamefont {Kim}},
  \bibinfo {author} {\bibfnamefont {Jiawei}\ \bibnamefont {Zang}}, \bibinfo
  {author} {\bibfnamefont {Andrew~J.}\ \bibnamefont {Millis}}, \bibinfo
  {author} {\bibfnamefont {Kenji}\ \bibnamefont {Watanabe}}, \bibinfo {author}
  {\bibfnamefont {Takashi}\ \bibnamefont {Taniguchi}}, \bibinfo {author}
  {\bibfnamefont {James~C.}\ \bibnamefont {Hone}}, \bibinfo {author}
  {\bibfnamefont {Lei}\ \bibnamefont {Wang}}, \bibinfo {author} {\bibfnamefont
  {Cory~R.}\ \bibnamefont {Dean}}, \ and\ \bibinfo {author} {\bibfnamefont
  {Abhay~N.}\ \bibnamefont {Pasupathy}},\ }\bibfield  {title} {\enquote
  {\bibinfo {title} {Quantum criticality in twisted transition metal
  dichalcogenides},}\ }\href {\doibase 10.1038/s41586-021-03815-6} {\bibfield
  {journal} {\bibinfo  {journal} {Nature}\ }\textbf {\bibinfo {volume} {597}},\
  \bibinfo {pages} {345--349} (\bibinfo {year} {2021})}\BibitemShut {NoStop}%
\bibitem [{\citenamefont {Coppersmith}\ \emph {et~al.}(1982)\citenamefont
  {Coppersmith}, \citenamefont {Fisher}, \citenamefont {Halperin},
  \citenamefont {Lee},\ and\ \citenamefont
  {Brinkman}}]{Coppersmith1982Phys.Rev.B}%
  \BibitemOpen
  \bibfield  {author} {\bibinfo {author} {\bibfnamefont {S.~N.}\ \bibnamefont
  {Coppersmith}}, \bibinfo {author} {\bibfnamefont {Daniel~S.}\ \bibnamefont
  {Fisher}}, \bibinfo {author} {\bibfnamefont {B.~I.}\ \bibnamefont
  {Halperin}}, \bibinfo {author} {\bibfnamefont {P.~A.}\ \bibnamefont {Lee}}, \
  and\ \bibinfo {author} {\bibfnamefont {W.~F.}\ \bibnamefont {Brinkman}},\
  }\bibfield  {title} {\enquote {\bibinfo {title} {Dislocations and the
  commensurate-incommensurate transition in two dimensions},}\ }\href {\doibase
  10.1103/PhysRevB.25.349} {\bibfield  {journal} {\bibinfo  {journal} {Phys.
  Rev. B}\ }\textbf {\bibinfo {volume} {25}},\ \bibinfo {pages} {349--363}
  (\bibinfo {year} {1982})}\BibitemShut {NoStop}%
\bibitem [{\citenamefont {Kivelson}\ \emph {et~al.}(1998)\citenamefont
  {Kivelson}, \citenamefont {Fradkin},\ and\ \citenamefont
  {Emery}}]{Kivelson1998Nature}%
  \BibitemOpen
  \bibfield  {author} {\bibinfo {author} {\bibfnamefont {S.~A.}\ \bibnamefont
  {Kivelson}}, \bibinfo {author} {\bibfnamefont {E.}~\bibnamefont {Fradkin}}, \
  and\ \bibinfo {author} {\bibfnamefont {V.~J.}\ \bibnamefont {Emery}},\
  }\bibfield  {title} {\enquote {\bibinfo {title} {Electronic liquid-crystal
  phases of a doped {{Mott}} insulator},}\ }\href {\doibase 10.1038/31177}
  {\bibfield  {journal} {\bibinfo  {journal} {Nature}\ }\textbf {\bibinfo
  {volume} {393}},\ \bibinfo {pages} {550--553} (\bibinfo {year}
  {1998})}\BibitemShut {NoStop}%
\bibitem [{\citenamefont {Serre}(2012)}]{Serre2012}%
  \BibitemOpen
  \bibfield  {author} {\bibinfo {author} {\bibfnamefont {Jean-Pierre}\
  \bibnamefont {Serre}},\ }\href@noop {} {\emph {\bibinfo {title} {Linear
  {{Representations}} of {{Finite Groups}}}}}\ (\bibinfo  {publisher}
  {{Springer Science \& Business Media}},\ \bibinfo {year} {2012})\BibitemShut
  {NoStop}%
\bibitem [{\citenamefont {Fernandes}\ and\ \citenamefont
  {Venderbos}(2020)}]{Fernandes2020Sci.Adv.}%
  \BibitemOpen
  \bibfield  {author} {\bibinfo {author} {\bibfnamefont {Rafael~M.}\
  \bibnamefont {Fernandes}}\ and\ \bibinfo {author} {\bibfnamefont {J{\"o}rn
  W.~F.}\ \bibnamefont {Venderbos}},\ }\bibfield  {title} {\enquote {\bibinfo
  {title} {Nematicity with a twist: {{Rotational}} symmetry breaking in a
  moir\'e superlattice},}\ }\href {\doibase 10.1126/sciadv.aba8834} {\bibfield
  {journal} {\bibinfo  {journal} {Science Advances}\ }\textbf {\bibinfo
  {volume} {6}} (\bibinfo {year} {2020}),\ 10.1126/sciadv.aba8834},\ \Eprint
  {http://arxiv.org/abs/https://advances.sciencemag.org/content/6/32/eaba8834.full.pdf}
  {https://advances.sciencemag.org/content/6/32/eaba8834.full.pdf} \BibitemShut
  {NoStop}%
\bibitem [{\citenamefont {Kivelson}\ \emph {et~al.}(2003)\citenamefont
  {Kivelson}, \citenamefont {Bindloss}, \citenamefont {Fradkin}, \citenamefont
  {Oganesyan}, \citenamefont {Tranquada}, \citenamefont {Kapitulnik},\ and\
  \citenamefont {Howald}}]{Kivelson2003Rev.Mod.Phys.}%
  \BibitemOpen
  \bibfield  {author} {\bibinfo {author} {\bibfnamefont {S.~A.}\ \bibnamefont
  {Kivelson}}, \bibinfo {author} {\bibfnamefont {I.~P.}\ \bibnamefont
  {Bindloss}}, \bibinfo {author} {\bibfnamefont {E.}~\bibnamefont {Fradkin}},
  \bibinfo {author} {\bibfnamefont {V.}~\bibnamefont {Oganesyan}}, \bibinfo
  {author} {\bibfnamefont {J.~M.}\ \bibnamefont {Tranquada}}, \bibinfo {author}
  {\bibfnamefont {A.}~\bibnamefont {Kapitulnik}}, \ and\ \bibinfo {author}
  {\bibfnamefont {C.}~\bibnamefont {Howald}},\ }\bibfield  {title} {\enquote
  {\bibinfo {title} {How to detect fluctuating stripes in the high-temperature
  superconductors},}\ }\href {\doibase 10.1103/RevModPhys.75.1201} {\bibfield
  {journal} {\bibinfo  {journal} {Rev. Mod. Phys.}\ }\textbf {\bibinfo {volume}
  {75}},\ \bibinfo {pages} {1201--1241} (\bibinfo {year} {2003})}\BibitemShut
  {NoStop}%
\bibitem [{\citenamefont {Venderbos}\ and\ \citenamefont
  {Fernandes}(2018)}]{Venderbos2018Phys.Rev.B}%
  \BibitemOpen
  \bibfield  {author} {\bibinfo {author} {\bibfnamefont {J{\"o}rn W.~F.}\
  \bibnamefont {Venderbos}}\ and\ \bibinfo {author} {\bibfnamefont {Rafael~M.}\
  \bibnamefont {Fernandes}},\ }\bibfield  {title} {\enquote {\bibinfo {title}
  {Correlations and electronic order in a two-orbital honeycomb lattice model
  for twisted bilayer graphene},}\ }\href {\doibase 10.1103/PhysRevB.98.245103}
  {\bibfield  {journal} {\bibinfo  {journal} {Phys. Rev. B}\ }\textbf {\bibinfo
  {volume} {98}},\ \bibinfo {pages} {245103} (\bibinfo {year}
  {2018})}\BibitemShut {NoStop}%
\bibitem [{\citenamefont {Hecker}\ and\ \citenamefont
  {Schmalian}(2018)}]{Hecker2018npjQuantMater}%
  \BibitemOpen
  \bibfield  {author} {\bibinfo {author} {\bibfnamefont {Matthias}\
  \bibnamefont {Hecker}}\ and\ \bibinfo {author} {\bibfnamefont {J{\"o}rg}\
  \bibnamefont {Schmalian}},\ }\bibfield  {title} {\enquote {\bibinfo {title}
  {Vestigial nematic order and superconductivity in the doped topological
  insulator {{Cu}} x {{Bi2Se3}}},}\ }\href {\doibase 10.1038/s41535-018-0098-z}
  {\bibfield  {journal} {\bibinfo  {journal} {npj Quant Mater}\ }\textbf
  {\bibinfo {volume} {3}},\ \bibinfo {pages} {26} (\bibinfo {year}
  {2018})}\BibitemShut {NoStop}%
\bibitem [{\citenamefont {Little}\ \emph {et~al.}(2020)\citenamefont {Little},
  \citenamefont {Lee}, \citenamefont {John}, \citenamefont {Doyle},
  \citenamefont {Maniv}, \citenamefont {Nair}, \citenamefont {Chen},
  \citenamefont {Rees}, \citenamefont {Venderbos}, \citenamefont {Fernandes},
  \citenamefont {Analytis},\ and\ \citenamefont
  {Orenstein}}]{Little2020Nat.Mater.}%
  \BibitemOpen
  \bibfield  {author} {\bibinfo {author} {\bibfnamefont {Arielle}\ \bibnamefont
  {Little}}, \bibinfo {author} {\bibfnamefont {Changmin}\ \bibnamefont {Lee}},
  \bibinfo {author} {\bibfnamefont {Caolan}\ \bibnamefont {John}}, \bibinfo
  {author} {\bibfnamefont {Spencer}\ \bibnamefont {Doyle}}, \bibinfo {author}
  {\bibfnamefont {Eran}\ \bibnamefont {Maniv}}, \bibinfo {author}
  {\bibfnamefont {Nityan~L.}\ \bibnamefont {Nair}}, \bibinfo {author}
  {\bibfnamefont {Wenqin}\ \bibnamefont {Chen}}, \bibinfo {author}
  {\bibfnamefont {Dylan}\ \bibnamefont {Rees}}, \bibinfo {author}
  {\bibfnamefont {J{\"o}rn W.~F.}\ \bibnamefont {Venderbos}}, \bibinfo {author}
  {\bibfnamefont {Rafael~M.}\ \bibnamefont {Fernandes}}, \bibinfo {author}
  {\bibfnamefont {James~G.}\ \bibnamefont {Analytis}}, \ and\ \bibinfo {author}
  {\bibfnamefont {Joseph}\ \bibnamefont {Orenstein}},\ }\bibfield  {title}
  {\enquote {\bibinfo {title} {Three-state nematicity in the triangular lattice
  antiferromagnet {{Fe1}}/{{3NbS2}}},}\ }\href {\doibase
  10.1038/s41563-020-0681-0} {\bibfield  {journal} {\bibinfo  {journal} {Nat.
  Mater.}\ }\textbf {\bibinfo {volume} {19}},\ \bibinfo {pages} {1062--1067}
  (\bibinfo {year} {2020})}\BibitemShut {NoStop}%
\bibitem [{\citenamefont {Li}\ \emph {et~al.}(2021{\natexlab{c}})\citenamefont
  {Li}, \citenamefont {Li}, \citenamefont {Naik}, \citenamefont {Xie},
  \citenamefont {Li}, \citenamefont {Wang}, \citenamefont {Regan},
  \citenamefont {Wang}, \citenamefont {Zhao}, \citenamefont {Zhao},
  \citenamefont {Kahn}, \citenamefont {Yumigeta}, \citenamefont {Blei},
  \citenamefont {Taniguchi}, \citenamefont {Watanabe}, \citenamefont {Tongay},
  \citenamefont {Zettl}, \citenamefont {Louie}, \citenamefont {Wang},\ and\
  \citenamefont {Crommie}}]{Li2021Nat.Mater.}%
  \BibitemOpen
  \bibfield  {author} {\bibinfo {author} {\bibfnamefont {Hongyuan}\
  \bibnamefont {Li}}, \bibinfo {author} {\bibfnamefont {Shaowei}\ \bibnamefont
  {Li}}, \bibinfo {author} {\bibfnamefont {Mit~H.}\ \bibnamefont {Naik}},
  \bibinfo {author} {\bibfnamefont {Jingxu}\ \bibnamefont {Xie}}, \bibinfo
  {author} {\bibfnamefont {Xinyu}\ \bibnamefont {Li}}, \bibinfo {author}
  {\bibfnamefont {Jiayin}\ \bibnamefont {Wang}}, \bibinfo {author}
  {\bibfnamefont {Emma}\ \bibnamefont {Regan}}, \bibinfo {author}
  {\bibfnamefont {Danqing}\ \bibnamefont {Wang}}, \bibinfo {author}
  {\bibfnamefont {Wenyu}\ \bibnamefont {Zhao}}, \bibinfo {author}
  {\bibfnamefont {Sihan}\ \bibnamefont {Zhao}}, \bibinfo {author}
  {\bibfnamefont {Salman}\ \bibnamefont {Kahn}}, \bibinfo {author}
  {\bibfnamefont {Kentaro}\ \bibnamefont {Yumigeta}}, \bibinfo {author}
  {\bibfnamefont {Mark}\ \bibnamefont {Blei}}, \bibinfo {author} {\bibfnamefont
  {Takashi}\ \bibnamefont {Taniguchi}}, \bibinfo {author} {\bibfnamefont
  {Kenji}\ \bibnamefont {Watanabe}}, \bibinfo {author} {\bibfnamefont
  {Sefaattin}\ \bibnamefont {Tongay}}, \bibinfo {author} {\bibfnamefont {Alex}\
  \bibnamefont {Zettl}}, \bibinfo {author} {\bibfnamefont {Steven~G.}\
  \bibnamefont {Louie}}, \bibinfo {author} {\bibfnamefont {Feng}\ \bibnamefont
  {Wang}}, \ and\ \bibinfo {author} {\bibfnamefont {Michael~F.}\ \bibnamefont
  {Crommie}},\ }\bibfield  {title} {\enquote {\bibinfo {title} {Imaging
  moir\'eflat bands in three-dimensional reconstructed {{WSe2}}/{{WS2}}
  superlattices},}\ }\href {\doibase 10.1038/s41563-021-00923-6} {\bibfield
  {journal} {\bibinfo  {journal} {Nature Materials}\ } (\bibinfo {year}
  {2021}{\natexlab{c}}),\ 10.1038/s41563-021-00923-6}\BibitemShut {NoStop}%
\bibitem [{\citenamefont {Wolff}(1989)}]{Wolff1989Phys.Rev.Lett.}%
  \BibitemOpen
  \bibfield  {author} {\bibinfo {author} {\bibfnamefont {Ulli}\ \bibnamefont
  {Wolff}},\ }\bibfield  {title} {\enquote {\bibinfo {title} {Collective
  {{Monte Carlo Updating}} for {{Spin Systems}}},}\ }\href {\doibase
  10.1103/PhysRevLett.62.361} {\bibfield  {journal} {\bibinfo  {journal} {Phys.
  Rev. Lett.}\ }\textbf {\bibinfo {volume} {62}},\ \bibinfo {pages} {361--364}
  (\bibinfo {year} {1989})}\BibitemShut {NoStop}%
\bibitem [{\citenamefont {Heringa}\ and\ \citenamefont
  {Bl{\"o}te}(1998)}]{Heringa1998Phys.Rev.E}%
  \BibitemOpen
  \bibfield  {author} {\bibinfo {author} {\bibfnamefont {J.~R.}\ \bibnamefont
  {Heringa}}\ and\ \bibinfo {author} {\bibfnamefont {H.~W.~J.}\ \bibnamefont
  {Bl{\"o}te}},\ }\bibfield  {title} {\enquote {\bibinfo {title} {Geometric
  cluster {{Monte Carlo}} simulation},}\ }\href {\doibase
  10.1103/PhysRevE.57.4976} {\bibfield  {journal} {\bibinfo  {journal} {Phys.
  Rev. E}\ }\textbf {\bibinfo {volume} {57}},\ \bibinfo {pages} {4976--4978}
  (\bibinfo {year} {1998})}\BibitemShut {NoStop}%
\bibitem [{\citenamefont {Shimazaki}\ \emph {et~al.}(2021)\citenamefont
  {Shimazaki}, \citenamefont {Kuhlenkamp}, \citenamefont {Schwartz},
  \citenamefont {Smole{\'n}ski}, \citenamefont {Watanabe}, \citenamefont
  {Taniguchi}, \citenamefont {Kroner}, \citenamefont {Schmidt}, \citenamefont
  {Knap},\ and\ \citenamefont {Imamo{\u g}lu}}]{Shimazaki2021Phys.Rev.X}%
  \BibitemOpen
  \bibfield  {author} {\bibinfo {author} {\bibfnamefont {Yuya}\ \bibnamefont
  {Shimazaki}}, \bibinfo {author} {\bibfnamefont {Clemens}\ \bibnamefont
  {Kuhlenkamp}}, \bibinfo {author} {\bibfnamefont {Ido}\ \bibnamefont
  {Schwartz}}, \bibinfo {author} {\bibfnamefont {Tomasz}\ \bibnamefont
  {Smole{\'n}ski}}, \bibinfo {author} {\bibfnamefont {Kenji}\ \bibnamefont
  {Watanabe}}, \bibinfo {author} {\bibfnamefont {Takashi}\ \bibnamefont
  {Taniguchi}}, \bibinfo {author} {\bibfnamefont {Martin}\ \bibnamefont
  {Kroner}}, \bibinfo {author} {\bibfnamefont {Richard}\ \bibnamefont
  {Schmidt}}, \bibinfo {author} {\bibfnamefont {Michael}\ \bibnamefont {Knap}},
  \ and\ \bibinfo {author} {\bibfnamefont {Ata{\c c}}\ \bibnamefont {Imamo{\u
  g}lu}},\ }\bibfield  {title} {\enquote {\bibinfo {title} {Optical
  {{Signatures}} of {{Periodic Charge Distribution}} in a {{Mott-like
  Correlated Insulator State}}},}\ }\href {\doibase 10.1103/PhysRevX.11.021027}
  {\bibfield  {journal} {\bibinfo  {journal} {Phys. Rev. X}\ }\textbf {\bibinfo
  {volume} {11}},\ \bibinfo {pages} {021027} (\bibinfo {year}
  {2021})}\BibitemShut {NoStop}%
\bibitem [{\citenamefont {Zhou}\ \emph {et~al.}(2021)\citenamefont {Zhou},
  \citenamefont {Sung}, \citenamefont {Brutschea}, \citenamefont {Esterlis},
  \citenamefont {Wang}, \citenamefont {Scuri}, \citenamefont {Gelly},
  \citenamefont {Heo}, \citenamefont {Taniguchi}, \citenamefont {Watanabe},
  \citenamefont {Zar{\'a}nd}, \citenamefont {Lukin}, \citenamefont {Kim},
  \citenamefont {Demler},\ and\ \citenamefont {Park}}]{Zhou2021Nature}%
  \BibitemOpen
  \bibfield  {author} {\bibinfo {author} {\bibfnamefont {You}\ \bibnamefont
  {Zhou}}, \bibinfo {author} {\bibfnamefont {Jiho}\ \bibnamefont {Sung}},
  \bibinfo {author} {\bibfnamefont {Elise}\ \bibnamefont {Brutschea}}, \bibinfo
  {author} {\bibfnamefont {Ilya}\ \bibnamefont {Esterlis}}, \bibinfo {author}
  {\bibfnamefont {Yao}\ \bibnamefont {Wang}}, \bibinfo {author} {\bibfnamefont
  {Giovanni}\ \bibnamefont {Scuri}}, \bibinfo {author} {\bibfnamefont
  {Ryan~J.}\ \bibnamefont {Gelly}}, \bibinfo {author} {\bibfnamefont {Hoseok}\
  \bibnamefont {Heo}}, \bibinfo {author} {\bibfnamefont {Takashi}\ \bibnamefont
  {Taniguchi}}, \bibinfo {author} {\bibfnamefont {Kenji}\ \bibnamefont
  {Watanabe}}, \bibinfo {author} {\bibfnamefont {Gergely}\ \bibnamefont
  {Zar{\'a}nd}}, \bibinfo {author} {\bibfnamefont {Mikhail~D.}\ \bibnamefont
  {Lukin}}, \bibinfo {author} {\bibfnamefont {Philip}\ \bibnamefont {Kim}},
  \bibinfo {author} {\bibfnamefont {Eugene}\ \bibnamefont {Demler}}, \ and\
  \bibinfo {author} {\bibfnamefont {Hongkun}\ \bibnamefont {Park}},\ }\bibfield
   {title} {\enquote {\bibinfo {title} {Bilayer {{Wigner}} crystals in a
  transition metal dichalcogenide heterostructure},}\ }\href {\doibase
  10.1038/s41586-021-03560-w} {\bibfield  {journal} {\bibinfo  {journal}
  {Nature}\ }\textbf {\bibinfo {volume} {595}},\ \bibinfo {pages} {48--52}
  (\bibinfo {year} {2021})}\BibitemShut {NoStop}%
\bibitem [{\citenamefont {Smole{\'n}ski}\ \emph {et~al.}(2021)\citenamefont
  {Smole{\'n}ski}, \citenamefont {Dolgirev}, \citenamefont {Kuhlenkamp},
  \citenamefont {Popert}, \citenamefont {Shimazaki}, \citenamefont {Back},
  \citenamefont {Lu}, \citenamefont {Kroner}, \citenamefont {Watanabe},
  \citenamefont {Taniguchi}, \citenamefont {Esterlis}, \citenamefont {Demler},\
  and\ \citenamefont {Imamo{\u g}lu}}]{Smolenski2021Nature}%
  \BibitemOpen
  \bibfield  {author} {\bibinfo {author} {\bibfnamefont {Tomasz}\ \bibnamefont
  {Smole{\'n}ski}}, \bibinfo {author} {\bibfnamefont {Pavel~E.}\ \bibnamefont
  {Dolgirev}}, \bibinfo {author} {\bibfnamefont {Clemens}\ \bibnamefont
  {Kuhlenkamp}}, \bibinfo {author} {\bibfnamefont {Alexander}\ \bibnamefont
  {Popert}}, \bibinfo {author} {\bibfnamefont {Yuya}\ \bibnamefont
  {Shimazaki}}, \bibinfo {author} {\bibfnamefont {Patrick}\ \bibnamefont
  {Back}}, \bibinfo {author} {\bibfnamefont {Xiaobo}\ \bibnamefont {Lu}},
  \bibinfo {author} {\bibfnamefont {Martin}\ \bibnamefont {Kroner}}, \bibinfo
  {author} {\bibfnamefont {Kenji}\ \bibnamefont {Watanabe}}, \bibinfo {author}
  {\bibfnamefont {Takashi}\ \bibnamefont {Taniguchi}}, \bibinfo {author}
  {\bibfnamefont {Ilya}\ \bibnamefont {Esterlis}}, \bibinfo {author}
  {\bibfnamefont {Eugene}\ \bibnamefont {Demler}}, \ and\ \bibinfo {author}
  {\bibfnamefont {Ata{\c c}}\ \bibnamefont {Imamo{\u g}lu}},\ }\bibfield
  {title} {\enquote {\bibinfo {title} {Signatures of {{Wigner}} crystal of
  electrons in a monolayer semiconductor},}\ }\href {\doibase
  10.1038/s41586-021-03590-4} {\bibfield  {journal} {\bibinfo  {journal}
  {Nature}\ }\textbf {\bibinfo {volume} {595}},\ \bibinfo {pages} {53--57}
  (\bibinfo {year} {2021})}\BibitemShut {NoStop}%
\end{thebibliography}%

\appendix
\section{Appendix A: Cluster Algorithm for Monte Carlo Simulation}
This appendix details the cluster algorithm we developed for our Monte Carlo
simulations of the triangular Ising lattice gas with long range interactions.
This algorithm is similar in spirit to the well-known 
Wolff algorithm~\cite{Wolff1989Phys.Rev.Lett.}
and its later generalization: the so-called geometric cluster algorithm
used to simulate the fixed magnetization ensemble of the nearest-neighbor 
Ising model~\cite{Heringa1998Phys.Rev.E}.
It is useful to consider the
triangular lattice with $N$ sites as having its sites indexed by integer values
$i \in \mathbb{Z}_N$.
We define an injective
map $\mathcal{L} : \mathbb{Z}_N \rightarrow \mathbb{R}^2$ that takes an integer
lattice site index and maps it to a real space coordinate. The precise
action of this map depends on the details of the finite-size geometry. However, it 
will always return a linear combination of the triangular lattice vectors, i.e.
$\mathcal{L}(i) = f(i) \vec{a}_1 + g(i) \vec{a}_2$ with lattice unit vectors
$\vec{a}_{1,2}$ such that $\vec{a}_1 \cdot \vec{a}_2 = \pm a^2/2$
for some integer valued functions $f,g$ and where $a \in \mathbb{R}^+$ is the lattice constant.

One can view a particle configuration as a set $\Lambda \subseteq \mathbb{Z}_N$ specifying occupied
sites of the lattice. 
The Lattice is occupied by Ising variables, and the
occupancy function $n : \mathbb{Z}_N \times \mathbb{Z}_N \rightarrow \mathbb{Z}_2$ 
acts on a site index $i$ and configuration $\Lambda$ as
\[
    n(\Lambda,i) = \begin{cases} 1 & i \in \Lambda \\ 0 & \text{otherwise} \end{cases}.
\]
We are interested in the case of a fixed number of particles $M \leq N$, i.e. the only valid
particle configurations $\Lambda$ are those with cardinality $M$.

We treat elements of the triangular lattice point group $\tau \in D_3$ as maps
$\tau : \mathbb{Z}_N \rightarrow \mathbb{Z}_N$ which map lattice site indices to lattice
site indices. Particle exchange is a map 
$\eta: \mathbb{Z}_N \times \mathbb{Z}_N \times D_3 \rightarrow \mathbb{Z}_N$
defined as
\[
    \eta(\Lambda, i, \tau) = \begin{cases} 
    \Lambda & n(\Lambda, i) = n(\Lambda, \tau(i)) \\
    (\Lambda \setminus \{i\}) \cup \{\tau(i)\} & i\in \Lambda,\tau(i)\notin\Lambda  \\
    (\Lambda \setminus \{\tau(i)\}) \cup \{i\} & \text{otherwise}
    \end{cases}.
\]
Note that $\eta$ preserves the cardinality of $\Lambda$.

The Hamiltonian for the system is 
$\mathcal{H}(\Lambda) = \frac{1}{2}\sum\limits_{i \neq j} V_{ij} n(\Lambda,i) n(\Lambda,j)$ 
where $V_{ij} = V(\vert\mathcal{L}(i) - \mathcal{L}(j)\vert)$. \\

\noindent \textbf{Algorithm}:
\begin{enumerate}
    \item Fix a particle configuration $\Lambda$, and initialize an empty set 
    $\mathcal{C}$ (the cluster)
    \item Choose an order-2 element, $\tau^*$, of the lattice point group.
    \item Randomly choose a site $i$, set $\mathcal{C} = \mathcal{C}\cup\{i,\tau^*(i)\}$
    and $\Lambda' = \eta(\Lambda,i,\tau^*)$
    \item For each other site $k$ with $V_{ik} \neq 0$ and $k \notin \mathcal{C}$:
    \begin{enumerate}
        \item
        Calculate 
        \begin{align*}
            \Delta_{ik}(\Lambda) \equiv \frac{1}{2} &[n(\Lambda,i)-n(\Lambda,\tau^*(i))]\times \\
            &[n(\Lambda,k)-n(\Lambda,\tau^*(k))]
            [V_{\tau^*(i),k}-V_{i,k}].
        \end{align*}
        The form of $\Delta_{ik}$ is chosen to ensure detailed balance
        (more detail in the next section).
        \item With probability $\max\left(0, 1-e^{-\beta\Delta_{ik}(\Lambda)}\right)$,
        set $\mathcal{C} = \mathcal{C} \cup \{k, \tau^*(k)\}$,
        $\Lambda' = \eta(\Lambda',k,\tau^*)$, and 
        record $k$ in a stack data structure
    \end{enumerate}
    \item Pop an element $j$ from the stack and repeat step $4$ with $j$ playing the 
    role of $i$
    \item Repeat step $5$ until the stack is empty and then return the updated particle
    configuration $\Lambda'$.
\end{enumerate}

\noindent\textbf{Proof of Detailed Balance:}\\

To prove that this generates the correct equilibrium probability distribution
(Boltzmann distribution),
we need to show that the probability of a particle configuration $\Lambda$ moving
to a configuration $\Lambda'$, $\mathcal{P}(\Lambda \rightarrow \Lambda')$, satisfies
the detailed balance condition, i.e.
\begin{align}\label{eqn:balance}
    \frac{ \mathcal{P}(\Lambda \rightarrow \Lambda') }
    { \mathcal{P}(\Lambda' \rightarrow \Lambda) } = 
    e^{-\beta(\mathcal{H}(\Lambda') - \mathcal{H}(\Lambda))}.
\end{align}
Observe that, for a move corresponding to a cluster $\mathcal{C}$,
we can write
$\mathcal{P}(\Lambda \rightarrow \Lambda') = 
\mathcal{P}_{\text{in}}(\mathcal{C},\Lambda)
\mathcal{P}_{\text{out}}(\mathcal{C},\Lambda)$ 
where the first factor is the probability of forming the cluster containing the 
sites in $\mathcal{C}$ and the second is the probability that no sites in 
$\mathbb{Z}_N \setminus \mathcal{C}$ are included in $\mathcal{C}$. Similarly,
we write $\mathcal{P}(\Lambda'\rightarrow\Lambda) = 
\mathcal{P}_{\text{in}}(\mathcal{C},\Lambda')
\mathcal{P}_{\text{out}}(\mathcal{C},\Lambda')$.

Because (1) $\tau^*$ is order-2 and a symmetry of $\mathcal{H}$ and (2)
$\mathcal{P}_{\text{in}}(\mathcal{C},\Lambda)$ only depends on lattice sites
in $\mathcal{C}$, we have that
$\mathcal{P}_{\text{in}}(\mathcal{C},\Lambda) = 
\mathcal{P}_{\text{in}}(\mathcal{C},\Lambda')$.

By construction of the algorithm, we have (using standard probability rules) that
\[
    \mathcal{P}_{\text{out}}(\mathcal{C},\Lambda) = 
    \exp\left( -\beta \frac{1}{2}
    \sum\limits_{i \in \mathcal{C}}\sum\limits_{k \notin \mathcal{C}}
    \max(0,\Delta_{ik}(\Lambda))
    \right)
\]
and 
\[
    \mathcal{P}_{\text{out}}(\mathcal{C},\Lambda') = 
    \exp\left( -\beta \frac{1}{2}
    \sum\limits_{i \in \mathcal{C}}\sum\limits_{k \notin \mathcal{C}}
    \max(0,-\Delta_{ik}(\Lambda))
    \right).
\]
where in the last equation we have used the fact that for
$i \in \mathcal{C}$, $n(\Lambda',i) = n(\Lambda,\tau^*(i))$.
Thus we have that
\begin{align}
    \frac{  \mathcal{P}_{\text{out}}(\mathcal{C},\Lambda) }
    { \mathcal{P}_{\text{out}}(\mathcal{C},\Lambda') } = 
    \exp\left( -\beta \frac{1}{2}
    \sum\limits_{i \in \mathcal{C}}\sum\limits_{k \notin \mathcal{C}}
    \Delta_{ik}(\Lambda)\right) \notag\\
    \label{eqn:lhs}
    = \exp\left( -\beta\frac{1}{2}\sum\limits_{i \in \mathcal{C}}\sum\limits_{k \notin \mathcal{C}}
    n(\Lambda,i)n(\Lambda,k)[V_{k,\tau^*(i)} - V_{k,i}]
    \right)
\end{align}
where in the second equality we have used that $\tau^*$ is order-2 and a symmetry of 
the Hamiltonian. Note also the factor of $1/2$, which is due to the fact that since
$i \in \mathcal{C} \Leftrightarrow \tau^*(i)\in\mathcal{C}$ and
$\Delta_{ik}(\Lambda) = \Delta_{\tau^*(i)k}(\Lambda)$ the sum
double counts. 

What remains is to show that the RHS of eq.~\ref{eqn:balance} is the same as
eq.~\ref{eqn:lhs}.
Using the fact that $\tau^*$ is order-2 and a symmetry of the Hamiltonian, we can write
\begin{align*}
    \mathcal{H}(\Lambda') &= \frac{1}{2}\sum\limits_{x \neq y, x,y \notin \mathcal{C}}
    n(\Lambda,x)n(\Lambda,y)V_{x,y}  \\
    &+\frac{1}{2}\sum\limits_{i \neq j, i,j \in\mathcal{C}}
    n(\Lambda,i)n(\Lambda,j)V_{i,j} \\ 
    &+\sum\limits_{i \in \mathcal{C}} \sum\limits_{x \notin \mathcal{C}}
    n(\Lambda,x)n(\Lambda,i)V_{x,\tau(i)}.
\end{align*}

It is then easy to see that the difference
\begin{align*}
    \mathcal{H}(\Lambda')-\mathcal{H}(\Lambda) &=
    \sum\limits_{i \in \mathcal{C}} \sum\limits_{x \notin \mathcal{C}}
    n(\Lambda,x)n(\Lambda,i)[V_{x,\tau^*(i)} - V_{x,i}].
\end{align*}
This completes the proof and establishes that the algorithm defined above satisfies detailed balance.\\

\noindent\textbf{A Note about Ergodicity:}\\
Note that by choosing $\tau^*$ as an appropriate reflection, it is 
possible to form a cluster consisting of a nearest neighbor particle exchange
with finite probability. Therefore, it is possible with finite probability
to get from any particle configuration to any other in the same way that
it is possible to realize any spin configuration in an Ising model via
a sequence of single spin flips.  Thus we can see that this algorithm is
indeed ergodic.

\section{Appendix B: Calculation of nematic correlation function and orientation }
To calculate the nematic order parameter expectation value
$\langle N(\vec{r})\rangle = \langle n(\vec{r})e^{i2\theta(\vec{r})} \rangle$
in Monte Carlo, we write $N(\vec{r})$ as
\begin{align}
    \vec{N}(\vec{r}) = \sum\limits_{\vec{\delta}} n_{\vec{r}} n_{\vec{r}+\vec{\delta}} 
    \begin{pmatrix}
    \delta_x^2 - \delta_y^2 \\ 2 \delta_x \delta_y
    \end{pmatrix}
\end{align}
where the vectors $\vec{\delta}$ point to the nearest neighbors of $\vec{r}$ and
$n_{\vec{r}}$ is the occupation number of the site at $\vec{r}$. This is easy to calculate
directly from a Monte Carlo configuration and average over the Markov chain. To calculate
the orientation $\langle \cos(6\theta) \rangle$ we use the fact that we can write the vector part of
the nematic order parameter as $(\cos(2\theta), \sin(2\theta))^T$. From this we
can simply calculate $\theta$ and hence $\cos(6\theta)$ from the nematic order parameter and average it
over the Markov chain.

\break
\section{Appendix C: Energetics of $\pi/3$ intersection and parallel dimer column fragments}
    We can understand why the $\pi/3$ intersections are energetically
    preferred schematically by calculating the energy 
    for two continuous, isolated wires of uniform line charge density $\lambda$ 
    interacting via eq.~\ref{eqn:hamiltonian} as a function of wire length $L$. 
    We approximate the wires as being at the center of the dimer columns, and plot the 
    results for parallel wires separated by $5a/2$ 
    in the purrple curve in Fig. S1.
    For a $\pi/3$ intersection with wires at the center of the dimers, the wires terminate
    with a separation of $\sqrt{3}a/2$. We show the results for such wires in the green
    curve in Fig. S1. Indeed, the energy of the parallel wires
    exceeds that of the angled wires for sufficient $L$. 
    
\begin{figure}
    \centering
    \includegraphics[width=\linewidth]{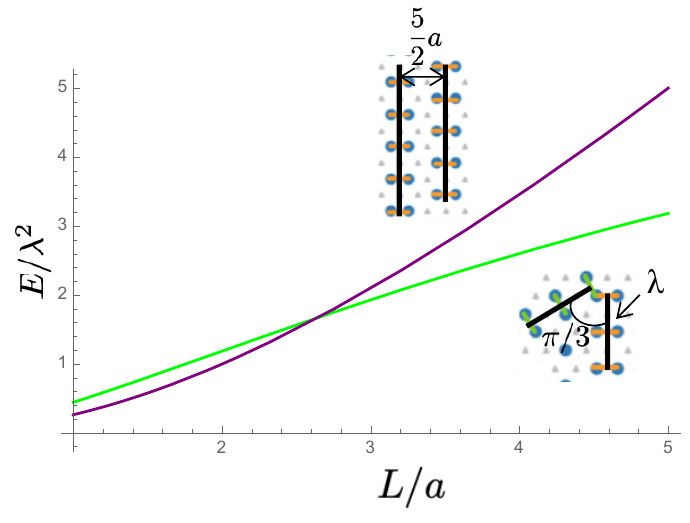}
    \caption{Energy of isolated, continuous wires interacting via eq.~\ref{eqn:hamiltonian} with 
    uniform line charge density $\lambda$ as a function of wire length $L$. We show results
    for parallel wires seperated by $5a/2$ (purple) and wires angled at $\pi/3$ with minimum
    separation $\sqrt{3}a/2$ (green). For sufficiently long wires the angled wires have lower energy.}
    \label{fig:wire_energy}
\end{figure}

\end{document}